\documentclass[twocolumn,showpacs,prb,citeautoscript,amsmath,amssymb,floatfix,eqsecnum]{revtex4-2}
\usepackage{amsmath,amssymb,color}
\usepackage{bm}
\usepackage{graphicx}
\usepackage{psfrag}
\usepackage{bbold}
\usepackage{bbm}
\newcommand{\bd}{\bm}
\definecolor{niklas}{RGB}{0,100,0}
\definecolor{max}{RGB}{0,0,220}

\begin{document}

\title{
Phonon renormalization and Pomeranchuk instability in the Holstein model}

\author{Niklas Cichutek}
\email[]{cichutek@itp.uni-frankfurt.de}
\author{Max Hansen}
\email[]{mhansen@itp.uni-frankfurt.de}
\author{Peter Kopietz}
  
\affiliation{Institut f\"{u}r Theoretische Physik, Universit\"{a}t
  Frankfurt,  Max-von-Laue Strasse 1, 60438 Frankfurt, Germany}

 \date{March 16, 2022}
%\date{December 6, 2008}

 \begin{abstract}
The Holstein model with dispersionless Einstein phonons is one of the simplest models
describing electron-phonon interactions in condensed matter. 
A naive  extrapolation of perturbation theory 
in powers of the relevant dimensionless electron-phonon coupling $\lambda_0$
suggests that at zero temperature the model exhibits 
a Pomeranchuk instability characterized by a divergent uniform compressibility
at a critical value of $\lambda_0$ of order unity.
In this work, we re-examine this problem
using modern functional renormalization group (RG) methods.
For dimensions $d > 3$ we find that
the RG flow of the Holstein model
indeed exhibits a tricritical fixed point
associated with a Pomeranchuk instability.
This non-Gaussian fixed point is ultraviolet stable and is closely related to
the well-known ultraviolet stable fixed point
of  $\phi^3$-theory above six dimensions.
To realize the Pomeranchuk critical point in the Holstein model
at fixed density both the electron-phonon coupling
$\lambda_0$ and the adiabatic ratio $\omega_0 / \epsilon_F$ 
have to be fine-tuned to assume critical values of order unity, where
$\omega_0$ is the phonon frequency and $\epsilon_F$ is the Fermi energy.
However, for dimensions $d \leq 3$ we find that the RG flow of the Holstein model does not have
any critical fixed points. 
This rules out a quantum critical point associated with a Pomeranchuk instability in $d \leq 3$.

\end{abstract}

%\pacs{XXXX}

\maketitle

\section{Introduction}

Gaining a microscopic understanding of the effect of
electron-phonon interactions on the physical behavior of metals continues to be 
one of the central
topics  in condensed matter physics. 
Although model systems for electron-phonon interactions have 
been studied for many decades using the established machinery of
many-body and kinetic theory \cite{Migdal58,Abrikosov88,Mahan00},
the limit of strong electron-phonon interactions remains a challenge to theory.
It is therefore not surprising that recently numerical calculations have 
revealed new phenomena in coupled electron-phonon systems which were not anticipated in the older 
literature, such as the emergence of
nontrivial phases with broken symmetry \cite{Kumar08,Murakami14,Ohgoe17,Esterlis18,Esterlis19,Chubukov20,Wang20} or the existence  of several hydrodynamic regimes with distinct
temperature dependence of various physical quantities~\cite{Huang21}.
Let us also point out that a microscopic description of
strongly coupled electron-phonon systems in the
so-called semiquantum regime \cite{Andreev78,Andreev79,Spivak10} 
$ \epsilon_F  \ll T \ll \omega_D  \ll V_{\rm cb}$
is still lacking.
Here $\epsilon_F$ is the Fermi energy, $T$ is the temperature, $\omega_D$ is the Debye energy, and $V_{\rm cb} \propto n^{1/3} e^2 $ is the characteristic Coulomb energy 
(where $n$ is the electronic density and $-e$ is the electric charge).
For simplicity, we measure temperature and frequencies in units of energy, which
amounts to formally setting $k_B = \hbar =1$. 
 
Given the fact that our understanding of strongly coupled 
electron-phonon systems is still incomplete, it is useful to approach this problem using
complementary methods. While numerical methods have produced a number of interesting
results, especially for the
phase diagram of coupled electron-phonon systems in  two dimensions~\cite{Kumar08,Murakami14,Ohgoe17,Esterlis18,Esterlis19,Chubukov20,Wang20}, a deeper understanding of the nature of phase transitions 
can be gained using  renormalization group (RG) methods.
Here we study 
the Holstein model \cite{Holstein59} in arbitrary dimensions $d$ using the functional 
renormalization group~\cite{Wetterich93,Berges02,Pawlowski07,Kopietz10,Metzner12,Dupuis21}. The Holstein model can be obtained as 
a special case of the Fr\"{o}hlich model \cite{Froehlich52} by assuming
a local electron-phonon coupling and  
dispersionless  Einstein phonons with frequency $\omega_0$.
Moreover, the direct  Coulomb interaction between the electrons is neglected.
In momentum space the Hamiltonian of the Holstein model is
 \begin{eqnarray}
   {\cal{H}} & = & \sum_{\bd{k}} \epsilon_{\bd{k}} c^{\dagger}_{\bd{k}} c_{\bd{k}} + \omega_0
 \sum_{\bd{q}} b^{\dagger}_{\bd{q}} b_{\bd{q}} 
% \nonumber
% \\
% & + & 
+ \frac{\gamma_0}{\sqrt{\cal{V}}}  
\sum_{\bd{k}, \bd{q} }  c^{\dagger}_{\bd{k} + \bd{q}}
 c_{\bd{k}}   
X_{\bd{q}},
 \hspace{7mm}
%\frac{b_{\bd{q}} + b^{\dagger}_{- \bd{q}}}{ \sqrt{2 \omega_0}} ,
 \label{eq:Hmigdal}
 \end{eqnarray}
where the operator
$c_{\bd{k}}$ annihilates an electron  with momentum $\bd{k}$ and
energy $\epsilon_{\bd{k}}$,
the operator  $b_{\bd{q}}$ annihilates a phonon with momentum
$\bd{q}$ and energy $\omega_0$, and
$X_{\bd{q}} = ( b_{\bd{q}} + b^{\dagger}_{- \bd{q}} )/
 \sqrt{2 \omega_0}$ is the Fourier component of the
phonon displacement  operator.
Here ${\cal{V}}$ is the volume of the system, 
$\bd{k}$-sums are over the first Brillouin zone,
$\bd{q}$-sums have an implicit ultraviolet cutoff of the order of the Debye momentum,
and we consider spinless electrons for simplicity.
The strength of the electron-phonon coupling
is characterized by the dimensionless parameter
 \begin{equation}
 \lambda_0 = \nu \gamma_0^2 / \omega_0^2,
 \end{equation}
where 
 \begin{equation}
\nu = \frac{1}{\cal{V}} \sum_{\bd{k}} \delta ( \epsilon_F
 - \epsilon_{\bd{k}} )
 \end{equation}
is the electronic density of states at the Fermi energy.

According to
Migdal's theorem \cite{Migdal58},   many-body calculations  for the Holstein model simplify in the regime
\begin{equation}
 \lambda_0 \omega_0 / \epsilon_F \ll  1,
 \label{eq:smalladiabatic}
 \end{equation}
because then vertex corrections can be neglected.
Note that in the adiabatic limit $ \omega_0 \ll \epsilon_F$ the
condition (\ref{eq:smalladiabatic}) is satisfied even for large values of $\lambda_0$, corresponding to strong electron-phonon interactions.
Recently, this generally accepted scenario  has been challenged 
for the two-dimensional Holstein model where, according to 
Refs.~[\onlinecite{Esterlis18,Esterlis19,Chubukov20}],   
nonperturbative effects become important when
$\lambda_0$ is of order order unity.
The breakdown of Migdal's theorem 
for the two-dimensional Holstein model
for $\lambda_0 = {\cal{O}} (1)$ 
is related to the nonperturbative formation of bipolarons \cite{Alexandrov81a,Alexandrov81b}
in a certain regime of temperatures.
In fact, a hint for the breakdown of perturbation theory 
in the Holstein model for 
$\lambda_0 = {\cal{O}} ( 1 )$ can already be obtained  by calculating the square of the
renormalized phonon frequency to first order in  $\lambda_0$, which yields \cite{Froehlich52,Sadovskii21}
 \begin{equation}
 \tilde{\omega}_0^2 = \omega_0^2 [ 1 - \lambda_0 + {\cal{O}} ( \lambda_0^2 ) ].
 \label{eq:phonren1}
 \end{equation}
If we boldly  neglect the higher-order corrections and extrapolate the first-order term in
Eq.~(\ref{eq:phonren1}) to intermediate coupling, then we find that
the renormalized phonon frequency vanishes for $\lambda_0 =1$.
But the renormalized phonon frequency of the Holstein model is related to 
the uniform  compressibility
$\partial \rho / \partial \mu$
via the Ward identity (see Appendix~A)
\begin{equation}
\tilde{\omega}_0^2 = \frac{\omega_0^2}{
 1 + \frac{ \gamma_0^2}{\omega_0^2} \frac{\partial \rho}{\partial \mu} },
 \label{eq:wiphon}
 \end{equation}
where $\rho$ is the electronic density and $\mu$ is the chemical potential.
The vanishing of $\tilde{\omega}_0^2$ for $\lambda_0 \rightarrow 1$ is therefore
accompanied by a divergence of the compressibility. 
At this point the system exhibits a Pomeranchuk instability 
in the zero angular momentum density channel associated with
phase separation \cite{Chubukov18}. Usually
Pomeranchuk instabilities in Fermi systems result from  
an effective electron-electron interaction~\cite{Chubukov18,Quintanilla06,Maslov10,Yamase09,Sarkar18,Quintanilla08}. 
Although the Holstein model does not include a direct interaction between the fermions,
by integrating over the phonons we can map the Holstein model onto an effective fermion model with a 
retarded two-body interaction, see Eq.~(\ref{eq:Seffc}) below.
The Pomeranchuk instability in the Holstein model can therefore be 
interpreted  conventionally  as an instability
of the Fermi liquid triggered by strong electron-electron interactions.
We will come back to this interpretation in Sec.~\ref{sec:FRGphononRGflow}, where
we also discuss the phase diagram which we show in Fig.~\ref{fig:Phasediagram}.
Historically, the vanishing of the perturbatively renormalized phonon frequency
for strong electron-phonon coupling
has already been noticed by Fr\"{o}hlich \cite{Froehlich52}, but
the question whether 
this  is a physical 
property of the Holstein model, or
an artefact of  the extrapolation of the first-order result (\ref{eq:phonren1}) has 
never been clarified.
Although some authors believe 
that in three dimensions the Holstein model
does not exhibit a Pomeranchuk instability \cite{Sadovskii21},
we have not been able to find a thorough investigation of this point in the literature.

In this work we  re-examine this problem using
modern FRG 
methods \cite{Wetterich93,Berges02,Pawlowski07,Kopietz10,Metzner12,Dupuis21}.
Our main result is that 
for $d > 3$ the RG flow of the Holstein model indeed exhibits
tricritical, ultraviolet-stable fixed point  
which can be associated with a quantum critical point where the system
exhibits a Pomeranchuk instability.
However, for dimensions   $d \leq  3$ the RG flow of the Holstein model
does not have any nontrivial fixed points but exhibits a runaway flow, which rules out
a Pomeranchuk instability characterized  
by a divergent uniform compressibility. While the interpretation of the
runaway RG flow at some finite scale is not unique, in Sec.~\ref{sec:FRGphononRGflow}
we argue that it indicates either a charge-density wave instability  or a first-order transition 
into an inhomogeneous state characterized by phase separation.

\section{Pomeranchuk instability from perturbation theory}
 \label{sec:pert}

\subsection{Calculation to second order in $\lambda_0$}

\begin{figure}
\begin{center}
  \centering
\vspace{7mm}
 \includegraphics[width=0.45\textwidth]{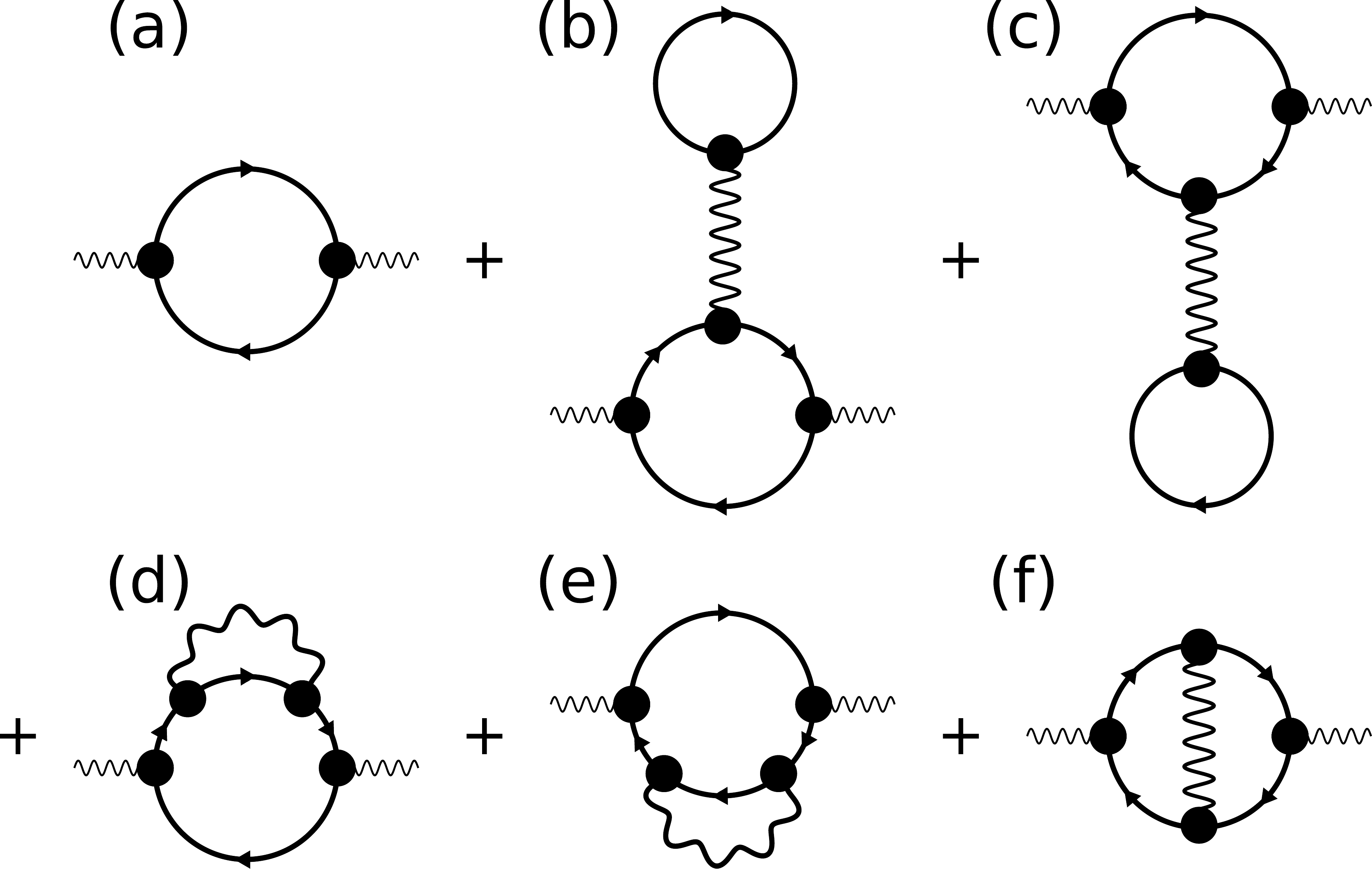}%\nspace
   \end{center}
%  \vspace{-4mm}
  \caption{%
%(Color online)
Perturbation series for the phonon self-energy $\Delta ( Q )$ in the 
Holstein model. Wavy lines represent the phonon propagator $D_0 ( Q )$, solid 
arrows  represent the electron propagator $G_0 ( K )$, and black dots represent the
bare electron-phonon vertex $\gamma_{{0}} \propto \sqrt{\lambda_0}$.
Diagram (a) represents the leading-order contribution 
$- \gamma_0^2 \Pi_0 ( Q )$
to the phonon self-energy, diagrams (b) and (c) are generated by  the Hartree corrections to the electronic self-energy, the diagrams (d) and (e) are due to  the
exchange corrections to the electronic self-energy, and diagram (f) represents the
leading vertex correction.
}
\label{fig:pertself}
\end{figure}

If we ignore the corrections of order $\lambda_0^2$ in
the perturbative expansion (\ref{eq:phonren1}) of the renormalized phonon frequency,
we find that the phonons soften for $\lambda_0 =1$. This value is of course not reliable and it is possible that the true renormalized phonon frequency $\tilde{\omega}_0 ( \lambda_0)$ never vanishes, even for large values of $\lambda_0$. 
As a first step in our investigation of this possibility, let us explicitly calculate the
correction of order $\lambda_0^2$ to the renormalized phonon frequency $\tilde{\omega}_0^2$.
Therefore we start from the Euclidean action
of the Holstein model (\ref{eq:Hmigdal}) and integrate over the 
momentum conjugate to the phonon displacement
$X_{Q}$ to arrive at the action
 \begin{eqnarray}
 S [  \bar{c} , c, X ] & = &  - \int_K {G}_0^{-1} ( K ) \bar{c}_K c_K + \frac{1}{2} \int_Q D_0^{-1} ( Q ) X_{-Q} X_Q
 \nonumber
  \\
 &  & + \gamma_0 \int_K \int_Q  \bar{c}_{ K+Q} c_K X_Q,
 \label{eq:ScX}
 \end{eqnarray}
where $c_K$ and $\bar{c}_K$ are Grassmann fields, the real bosonic field $X_Q$ 
represents the phonon displacement,
and the inverse noninteracting electron and phonon propagators are given by
 \begin{eqnarray}
%\bar{G}^{-1}_0 ( K ) =  - 
G_0^{-1} ( K )   & = & i \omega - \epsilon_{\bd{k}} +  \mu   ,
 \\
 D_0^{-1} ( Q )  & = &     \bar{\omega}^2 + \omega_0^2 .
 \end{eqnarray}
For convenience we have introduced collective labels $K = ( \bd{k} , i \omega )$ and
$Q = (  \bd{q} , i \bar{\omega} )$, where $i \omega$ are  
fermionic Matsubara frequencies and
$i \bar{\omega}$ are bosonic ones. The integration symbols are defined by
$\int_K = \frac{1}{ { \beta \cal{V}}  } \sum_{\bd{k}} \sum_{\omega}$ and
$\int_Q = \frac{1}{ \beta {\cal{V}}  } \sum_{\bd{q}} \sum_{\bar{\omega}}$,
where $\beta = 1/T$ is the inverse temperature.
The inverse propagators of the interacting system are of the form 
 \begin{eqnarray}
%\bar{G}^{-1}_0 ( K ) =  - 
G^{-1} ( K )   & = & i \omega - \epsilon_{\bd{k}} +  \mu  - \Sigma ( K )  ,
 \\
 D^{-1} ( Q )  & = &      \bar{\omega}^2  + \omega_0^2 + \Delta ( Q ),
 \end{eqnarray}
where $\Sigma ( K )$ is the electronic self-energy and the phonon self-energy $\Delta ( Q )$ 
can be expressed in terms of the interaction-irreducible polarization $\Pi (Q)$
as [see Eq.~(\ref{eq:DeltaPiconnection}) in Appendix~A]
 \begin{equation} 
 \Delta(Q)=-\gamma_0^2\Pi(Q).
 \end{equation}
The square of the renormalized phonon frequency for vanishing 
momentum is given by
 \begin{equation}
    \tilde{\omega}_0^2 = \omega_0^2 - \gamma_0^2 \lim_{\bd{q} \rightarrow 0} \Pi ( \bd{q} ,0 ).
 \end{equation}
To evaluate the $\mathcal{O}(\lambda_0^2)$ contribution to the renormalized squared
phonon frequency $\tilde{\omega}^2_0$, we consider the next-to-leading order contributions
to the irreducible polarization $\Pi(Q)$ represented  by the diagrams (b)--(f) 
in Fig.~\ref{fig:pertself}. We  can, analytically, perform all Matsubara sums and reduce the
evaluation of the diagrams, at $T=0$, to a two-dimensional integration over the energy variables $\xi$ and $\xi^{\prime}$. Our result for the $\bd{q}$-limit of the polarization is
 \begin{eqnarray}
 & & \Pi (0)  =   \lim_{\bd{q} \rightarrow 0} \Pi ( \bd{q} , 0 )   
 =  \nu  + \frac{\lambda_0}{3} \nu 
 \nonumber
 \\
 &  & +
 \frac{ \gamma_0^2}{2 \omega_0} \int_0^{\infty} d \xi
 \int_0^{\infty} d \xi^{\prime} 
 \Biggl[ \frac{ 4 \nu ( \mu + \xi ) \nu ( \mu - \xi^{\prime} )}{
 ( \xi + \xi^{\prime} + \omega_0 )^3} 
 \nonumber
 \\
 &  & +  \frac{ \nu ( \mu + \xi ) \nu^{\prime} ( \mu - \xi^{\prime} ) -
 \nu ( \mu - \xi ) \nu^{\prime} ( \mu + \xi^{\prime} ) }{
 ( \xi + \xi^{\prime} + \omega_0 )^2 }
 \Biggr]
% \nonumber
% \\
% & & 
+ {\cal{O}} ( \lambda_0^2 ),
 \nonumber
 \\
 & &
 \label{eq:Pi1}
 \end{eqnarray}
where $\nu ( \epsilon ) = \frac{1}{\cal{V}} \sum_{\bd{k}} \delta ( \epsilon - \epsilon_{\bd{k}} )$
is the energy-dependent density of states and
$\nu^{\prime} ( \epsilon ) = \partial \nu ( \epsilon ) / \partial \epsilon$.
The term $\lambda_0 \nu /3$ 
in the first line of Eq.~(\ref{eq:Pi1})
is due to the tadpole  self-energy corrections to the propagators 
in Fig.~\ref{fig:pertself} (b) and (c),
the term in the second line is due to the exchange self-energy diagrams (d) and (e),
and the last line involving the derivative of the density of states is due to the
vertex correction diagram (f).
In the adiabatic limit $\omega_0 \ll \epsilon_F$ the latter is 
a factor of $\omega_0 / \epsilon_F $ smaller than the self-energy contributions,
in accordance with Migdal's theorem. The double-integral in the second line 
evaluates to $2 \nu^2 / \omega_0$ so that for small $\omega_0 / \epsilon_F$ 
we obtain for the renormalized  phonon frequency
 \begin{equation}
 \tilde{\omega}_0^2 = \omega_0^2 \left[ 1 - \lambda_0 -  \frac{4}{3} \lambda_0^2 +
 {\cal{O}} ( \lambda_0^3 , \lambda_0^2 \omega_0 / \epsilon_F ) \right].
    \label{eq:Dinvexp}
 \end{equation} 
Neglecting the correction terms, we obtain an improved estimate
for the critical value $\lambda_c$ of the dimensionless electron-phonon 
coupling where the Pomeranchuk instability occurs,
 \begin{equation}
 \lambda_c = \frac{ \sqrt{57} -3}{8} \approx 0.57.
 \label{eq:lambdacres}
 \end{equation}
Surprisingly, the second-order correction reduces the
critical value of the electron-phonon coupling.
However, the critical $\lambda_c$ is still of the order of unity
so that the higher orders in  Eq.~(\ref{eq:Dinvexp})
cannot be neglected. In fact, we cannot exclude the  possibility that 
higher orders in $\lambda_0$ completely remove the
Pomeranchuk instability. 

\subsection{Effective phonon action and quantum critical point}

Since in this work we are only interested in the phonons, or rather the phonon self energy $\Delta(Q)$, it is
convenient to integrate over the fermions and work with the effective phonon
action. Thus, we may formally perform the Gaussian integration over the electron field
to obtain the effective phonon action
 \begin{equation}
 S_{\rm eff} [ X ] = \frac{1}{2} \int_Q D_0^{-1} ( Q ) X_{ - Q} X_Q - {\rm Tr} \ln 
 \left[ 1 - \gamma_0 \mathbf{G}_0 \mathbf{X} \right],
 \label{eq:Seffphonon}
 \end{equation} 
where $\mathbf{G}_0$ and $\mathbf{X}$ are infinite matrices in momentum-frequency space
with matrix elements
 \begin{eqnarray}
 [ \mathbf{G}_0 ]_{ K K^{\prime} } & = & \delta_{ K , K^{\prime}} G_0 ( K ),
 \\
 {[} \mathbf{X} {]}_{ K  K^{\prime}} & = & X_{ K - K^{\prime}},
 \end{eqnarray}
and the trace is normalized as follows,
 \begin{eqnarray}
 & &  - {\rm Tr} \ln 
 \left[ 1 - \gamma_0 \mathbf{G}_0 \mathbf{X} \right]  =  - \int_K 
 \ln 
 \left[ 1 - \gamma_0 \mathbf{G}_0 \mathbf{X} \right]_{ K K }
 \nonumber
 \\
 & = &  \gamma_0 \int_K  [ \mathbf{G}_0  \mathbf{X} ]_{ K K } + 
 \frac{\gamma_0^2}{2} \int_K  [  \mathbf{G}_0  \mathbf{X}   \mathbf{G}_0  \mathbf{X}    ]_{ K K } + \ldots
 \nonumber
 \\
 & = &  \gamma_0 X_0 \int_K G_0 ( K ) 
 \nonumber
 \\
 & &
+  \frac{\gamma_0^2}{2} \int_K \int_{K^{\prime}} 
  G_0 ( K ) X_{ K -K^{\prime}} G_0 ( K^{\prime} )  X_{ K^{\prime} - K } +
 \ldots \; .
 \nonumber
 \\
 & &
 \label{eq:tracelogexp}
 \end{eqnarray}
The expansion of the
effective action $S_{\rm eff} [ X ]$ in powers of the phonon field is therefore of the form
 \begin{eqnarray}
 & & S_{\rm eff} [ X ]  =  \Gamma_0^{(1)} X_{Q=0} + \frac{1}{2} \int_Q \Gamma_0^{(2)} ( Q ) 
 X_{-Q} X_Q 
 \nonumber
 \\
 &  & + \sum_{ n =3}^{\infty} \frac{1}{n!} \int_{Q_1} \cdots \int_{Q_n}
 \delta ( Q_1 + \ldots + Q_n ) 
 \nonumber
 \\
 & & \hspace{10mm} \times \Gamma_0^{(n)} ( Q_1 , \ldots , Q_n )
 X_{Q_1} \cdots X_{Q_n},
 \label{eq:SeffX}
 \end{eqnarray}
where
 \begin{eqnarray}
 \Gamma_0^{(1)} & = & \gamma_0 \int_K G_0 ( K ),
 \\
 \Gamma_0^{(2)} ( Q ) & = & D_0^{-1} ( Q ) + \gamma_0^{2} \int_K G_0 ( K ) G_0 ( K + Q ),
 \label{eq:Gammanull2}
 \hspace{7mm}
 \end{eqnarray}
and the interaction vertices can be expressed in terms of the symmetrized closed fermion loops $L_S^{(n)} ( Q_1 , \ldots , Q_n )$, defined in  Appendix~B, as follows,
 \begin{equation}
 \Gamma_0^{(n)} ( Q_1 , \ldots , Q_n ) = \gamma_0^n (n-1)! L_S^{(n)} ( -Q_1 , \ldots, - Q_n ).
 \label{eq:gammalsym}
 \end{equation}
To leading order in the electron-phonon interaction, the inverse phonon propagator is given by the 
function $\Gamma^{(2)}_0 ( Q )$
defined in Eq.~(\ref{eq:Gammanull2}) which can be written as
 \begin{equation}\label{eq:Dinvprop}
 D^{-1}_{\text{RPA}} ( Q ) = \Gamma_0^{(2)} ( Q ) =  \bar{\omega}^2 + \omega_0^2 -
 \gamma_0^2 \Pi_0 ( Q ),
 \end{equation}
 where we have introduced the noninteracting irreducible polarization
 \begin{eqnarray}
   \Pi_0 ( Q ) & = &  - \int_K G_0 ( K ) G_0 ( K + Q ) 
 \nonumber
 \\
 & = &  - \frac{1}{\cal{V}} \sum_{\bd{k}}
 \frac{ n_F ( \xi_{\bd{k} + \bd{q}} ) - n_F ( \xi_{\bd{k}} ) }{
 \epsilon_{\bd{k} + \bd{q}} - \epsilon_{\bd{k}} - i \bar{\omega} }.
 \end{eqnarray}
Here $\xi_{\bd{k}} = \epsilon_{\bd {k}} - \mu$ and $n_F ( \epsilon ) = 
 1/[ e^{  \epsilon /T  } +1 ]$ is the Fermi function. The approximation 
 (\ref{eq:Dinvprop}) is usually  
called random-phase approximation (RPA) and amounts to
neglecting  interaction corrections to the
irreducible polarization. 

The properties of $\Pi_0 ( Q )$ are well known in arbitrary dimensions \cite{Mihaila11}. For our purpose it is sufficient to expand $\Pi_0 ( Q )$ in the regime
$| \bar{\omega} | \ll v_F q$ and $ q \ll k_F$, where $k_F$ is the Fermi momentum.
At zero temperature the leading terms in the expansion are
 \begin{equation}
 \Pi_0 ( \bd{q} , i \bar{\omega} ) \approx 
\nu \left[ 1  - B_d \frac{ | \bar{\omega} | }{ v_F q } - C_d  \frac{q^2}{k_F^2}  \right],
 \label{eq:Piexpansion}
 \end{equation}
where the numerical coefficients $B_d$ and $C_d$ depend on the
dimensionality $d$ of the system and on the precise form of the dispersion $\epsilon_{\bd{k}}$. 
The coefficient $C_d$ is explicitly given by
 \begin{eqnarray} 
 C_d & = & \frac{ k_F^2}{8 \nu {\cal{V}}} \sum_{\bd{k}}
 \Bigl[
 n_F^{\prime \prime} ( \xi_{\bd{k} } )  ( \hat{\bd{q}} \cdot \bd{\nabla}_{\bd{k}} )^2 \epsilon_{\bd{k} }
 \nonumber
 \\
 & & \hspace{11mm}
 + \frac{1}{3}  n_F^{\prime \prime \prime} ( \xi_{\bd{k} } ) ( \hat{\bd{q}} \cdot \bd{\nabla}_{\bd{k}} \epsilon_{\bd{k}} )^2 \Bigr],
 \end{eqnarray}
where $n_F^{\prime \prime} ( \xi )$ and $n_F^{\prime \prime \prime} ( \xi )$ are the
second and third derivatives of the Fermi function, and $\hat{\bd{q}} = \bd{q} / q $ is a unit vector in the direction of $\bd{q}$.
In particular, for quadratic dispersion $\epsilon_{\bd{k}} = k^2 /(2m)$ at $T=0$
we have
 \begin{equation}
 B_3 = \frac{\pi}{2},  \; \; \; C_3 = \frac{1}{12}.
 \end{equation}
%
% \begin{equation}
% B_4 = \pi {\color{niklas} B_4 = 2},  \; \; \; C_4 = \frac{1}{6}.
% \end{equation}
% in three and four dimensions, respectively.
Inserting the expansion (\ref{eq:Piexpansion}) into the RPA propagator~(\ref{eq:Dinvprop}) we find that in this approximation the inverse
phonon propagator is
 \begin{eqnarray}
 D^{-1}_{\text{RPA}} ( Q ) 
 & \approx & r_0
 + b_0 | \bar{\omega} |  /q +   c_0  q^2  + {\cal{O}} ( \bar{\omega}^2 / q^2 , q^4 ) ,
 \label{eq:DinvRPA}
 \hspace{7mm}
 \end{eqnarray}
where
 \begin{subequations}
 \label{eq:rbcdef}
 \begin{eqnarray}
 r_0 & = & \tilde{\omega}_0^2 = \omega_0^2 ( 1 - \lambda_0 ),
 \label{eq:r0def}
 \\
b_0 & = &  B_d \gamma_0^2 \nu / v_F =
 B_d \lambda_0 \omega_0^2 /v_F,
 \label{eq:b0def}
 \\
 c_0 & = & C_d \gamma_0^2 \nu / k_F^2 =  C_d \lambda_0 \omega_0^2 / k_F^2.
 \label{eq:c0def}
 \end{eqnarray}
 \end{subequations}
The leading frequency dependence
 $B_d | \bar{\omega} | / ( v_F q )$ gives rise to Landau damping of the phonons, i.e., the decay of a single phonon into a 
fermionic particle-hole pair \cite{Pines89}.
The sign of the leading momentum dependence $C_d q^2 / k_F^2$ depends on the
dimensionality of the system;  for quadratic dispersion and at zero temperature $C_d $ is 
positive only for $d > 2$. For $d=2$ the coefficient $C_2 $ vanishes 
for quadratic dispersion because in this case the static 
Lindhard function $\Pi_0 ( \bd{q} , 0 )$ is 
constant in the interval $0 < q < 2k_F$ 
(see, for example, Ref.~[\onlinecite{Mihaila11}]). 

Within the RPA
the square of the renormalized phonon frequency in Eq.~(\ref{eq:r0def})
vanishes for $\lambda_0 =1$, as anticipated in Eq.~(\ref{eq:phonren1}).
However, the renormalized phonon frequency is related to the compressibility
 $\partial \rho / \partial \mu$ via the Ward identity (\ref{eq:wiphon}), which is closely related to the 
 compressibility sum rule
 \begin{equation}
 \frac{ \partial \rho}{\partial \mu} = \frac{ \Pi (0)}{ 1 - \frac{\gamma_0^2}{\omega_0^2} \Pi (0 ) },
 \end{equation}
as discussed in Appendix~A. The RPA result (\ref{eq:r0def}) for the renormalized
phonon frequency therefore implies that
for $\lambda_0 \rightarrow 1$ the compressibility diverges as
 \begin{equation}
 \frac{\partial \rho}{\partial \mu} = \frac{\nu}{ 1 - \lambda_0 },
 \label{eq:compdiv}
 \end{equation}
indicating a quantum critical point associated with a Pomeranchuk instability
in the zero angular momentum density channel \cite{Chubukov18}.
The perturbative result (\ref{eq:compdiv}) should be juxtaposed with the usual
Fermi-liquid expression of the compressibility \cite{Pines89}
 \begin{equation}
 \frac{\partial \rho}{\partial \mu} = \frac{\nu_{\ast}}{ 1 +F_0 },
 \label{eq:compFL}
 \end{equation}
 where $\nu_{\ast}$ is the renormalized density of states at the Fermi energy and $F_0$ is the Landau interaction parameter in the
zero angular momentum channel. Obviously,
$F_0 = - \lambda_0 < 0$ to leading order in perturbation theory.
In general, the Landau parameter
$F_0 ( \lambda_0 )$ is a complicated function of the bare coupling $\lambda_0$ and, naturally, the first-order result $F_0 = - \lambda_0$ cannot be trusted when
$\lambda_0$ is of order unity. Therefore, in the following section we will examine this problem 
in arbitrary dimensions using 
functional renormalization group methods, assuming that the 
zero angular momentum channel exhibits the dominant instability.
We do not examine the possibility of 
instabilities in higher angular momentum channels, which can in principle compete with the phase separation instability considered by us.

\section{Functional renormalization group approach}
\label{sec:FRGphonon}

From the Ward identity (\ref{eq:wiphon}) we see that a Pomeranchuk  quantum critical point 
is characterized by a divergent compressibility $ \partial \rho / \partial \mu$ and a vanishing renormalized
phonon frequency $\tilde{\omega}_0$. 
To investigate the possibility of a Pomeranchuk instability in  the
Holstein model, it is therefore sufficient to work with the effective phonon action
$S_{\rm eff} [ X ]$ defined in Eq.~(\ref{eq:Seffphonon}). We now use
the standard FRG machinery  \cite{Wetterich93,Berges02,Pawlowski07,Kopietz10,Metzner12,Dupuis21} 
to calculate 
the renormalized phonon frequency of  this effective field theory.

\subsection{Exact flow equations for the average effective phonon
action}

Following the usual procedure \cite{Wetterich93,Berges02,Pawlowski07,Kopietz10,Metzner12,Dupuis21} we now derive exact FRG flow equations for the
irreducible vertices of the effective field theory defined by the Euclidean action~(\ref{eq:Seffphonon}).
Therefore, we add a regulator to the bare action and consider 
 \begin{equation}
 S_{\Lambda} [ X ] = S_{\rm eff} [ X ] + \frac{1}{2} \int_Q 
 R_{\Lambda} ( Q ) X_{ - Q } X_Q,
 \end{equation}
where $R_{\Lambda} ( Q )$  introduces a scale parameter $\Lambda$ and
satisfies the boundary conditions
$R_{\Lambda =0} ( Q ) =0$ and $R_{\Lambda \rightarrow \infty} ( Q ) = \infty$.
We shall specify a convenient regulator in Sec.~\ref{sec:class}.
We then define the  scale-dependent 
average effective action $\Gamma_{\Lambda} [ \phi ]$ 
via the following subtracted Legendre transform of the generating functional
${\cal{G}}_{\Lambda} [ J ]$ of the  connected correlation functions~\cite{Wetterich93,Kopietz10},
 \begin{equation}
  \Gamma_{\Lambda} [ \phi ] = \int_Q \phi_{-Q} J_Q - {\cal{G}}_{\Lambda} [ J ] -
   \frac{1}{2} \int_Q R_{\Lambda} ( Q ) \phi_{ - Q } {\phi}_Q.
 \label{eq:averageactiondef}
 \end{equation}
Here the functional ${\cal{G}}_{\Lambda} [ J ]$ is defined by
 \begin{equation}
 e^{ {\cal{G}}_{\Lambda} [ J ] } = \int {\cal{D}} [ X ] e^{ - S_{\Lambda} [ X ] + \int_Q J_{-Q} X_Q },
 \end{equation} 
and the source field $J$ on the right-hand side of Eq.~(\ref{eq:averageactiondef})
should be expressed as functional of the field expectation values
$\phi_Q =  \langle X_Q \rangle$
by inverting the relation
 \begin{equation}
 \phi_Q = \frac{ \delta {\cal{G}}_{\Lambda} [ J ]}{\delta J_{-Q} }.
 \end{equation} 
The functional $\Gamma_{\Lambda} [ \phi ]$ satisfies the 
Wetterich equation \cite{Wetterich93}
\begin{equation}
 \partial_{\Lambda} \Gamma_{\Lambda} [ \phi ] = \frac{1}{2} {\rm Tr}  
\left[
 \left(
 \mathbf{\Gamma}^{\prime \prime }_{\Lambda} [ \phi ]  
 + \mathbf{R}_{\Lambda} \right)^{-1}
   \partial_{\Lambda} \mathbf{R}_{\Lambda} 
 \right],
 \label{eq:Wetterich1}
 \end{equation}
where $ \mathbf{\Gamma}^{\prime \prime }_{\Lambda} [ \phi ]  $
is the matrix of second functional derivatives of $\Gamma_{\Lambda} [ \phi ]$,
 \begin{equation}
 {\bigl[} 
\mathbf{\Gamma}^{\prime \prime }_{\Lambda} [ \phi ]  
{\bigr]}_{ Q Q^{\prime}} =
 \frac{ \delta^2 \Gamma_{\Lambda} [ \phi ] }{\delta \phi_{Q}
 \delta \phi_{Q^{\prime}  }},
 \end{equation}
and the elements of the 
regulator matrix $\mathbf{R}_{\Lambda}$ are
 \begin{equation}
 [{\mathbf{R}}_\Lambda ]_{ Q Q^{\prime}} = \delta ( Q + Q^{\prime} )  
 R_{\Lambda} ( Q^{\prime} ).
 \end{equation}
By construction, the functional $\Gamma_{\Lambda} [ \phi ]$ satisfies the boundary condition
 \begin{equation}
 \lim_{\Lambda_0 \rightarrow \infty} \Gamma_{\Lambda_0} [ \phi ] = S_{\rm eff} [ \phi ].
 \end{equation}

Due to the first-order  term $\Gamma_0^{(1)} X_{Q=0}$ in the bare action  (\ref{eq:SeffX})
the extremal condition for the average effective action
 \begin{equation}
 \frac{ \delta \Gamma_{\Lambda} [ \phi ]}{\delta {\phi}_Q } =0,
 \end{equation}
has a finite solution of the form ${\phi}^0_{\Lambda, Q} = \delta ( Q ) \phi^0_{\Lambda}$ with scale-dependent $\phi^0_{\Lambda}$ which for $\Lambda \rightarrow 0$ approaches the
value $\phi^0 = - \gamma_0 \rho / \omega_0^2$  given by
the Dyson-Schwinger equations derived in Appendix A, see 
Eqs.~(\ref{eq:DSphi0a}) and (\ref{eq:DSphi0}).
Usually, one would now set
  $
  {\phi}_Q = \phi^0_{\Lambda, Q} + \varphi_Q$
and consider the flow of the functional
  \begin{equation}
 \bar{\Gamma}_{\Lambda} [ \varphi ] = \Gamma_{\Lambda} [ \phi^0_{\Lambda} + \varphi ],
  \end{equation}
which satisfies the modified Wetterich equation
 \begin{eqnarray}
 \partial_{\Lambda} \bar{\Gamma}_{\Lambda} [ \varphi ] & = &
  \left. \partial_{\Lambda} {\Gamma}_{\Lambda} [ \phi ]
 \right|_{ \phi = \phi^0_{\Lambda} + \varphi }
% \nonumber
% \\
% & = & 
 +   \int_{Q} \frac{ \delta \bar{\Gamma}_{\Lambda} [ \varphi ]}{
 \delta \varphi_Q } \partial_{\Lambda} \phi^0_{\Lambda , Q}
 \nonumber
 \\
 & =  & \frac{1}{2} {\rm Tr}
\left[
 (
 \bar{\mathbf{\Gamma}}^{\prime \prime }_{\Lambda} [ \varphi ]  
 + \mathbf{R}_{\Lambda}  )^{-1}
   \partial_{\Lambda} \mathbf{R}_{\Lambda} 
 \right]
 \nonumber
 \\
    &  & 
+  \int_{Q} \frac{ \delta \bar{\Gamma}_{\Lambda} [ \varphi ]}{
 \delta \varphi_Q } \partial_{\Lambda} \phi^0_{\Lambda , Q}.
 \label{eq:Wetterichsym} 
 \end{eqnarray}
By construction
 \begin{equation}
 \left.
 \frac{ \delta \bar{\Gamma}_{\Lambda} [ \varphi ] }{\delta \varphi_Q }  \right|_{\varphi=0} =0,
 \end{equation}
so that
the vertex expansion of $\bar{\Gamma}_{\Lambda} [ \varphi ]$ in powers of $\varphi$ does not have a linear term.
However, 
this procedure implies that the electronic density is not held constant during the
RG flow, because according to Eq.~(\ref{eq:DSphi0}) the expectation values
$\phi^0_{\Lambda}$ is proportional to the density of electrons at scale $\Lambda$.
The resulting flow equations therefore relate  systems  with different electronic densities.
To keep the electronic density constant during the flow, 
we set $\phi = \phi^0 + \varphi$, where $\phi^0 = - \gamma_0 \rho / \omega_0^2$
is determined by the fixed (i.e., scale-independent) electronic density $\rho$.
We then set 
   \begin{equation}
  \tilde{\Gamma}_{\Lambda} [ \varphi ] = \Gamma_{\Lambda} [ \phi^0 + \varphi ],
  \end{equation}
which satisfies the usual Wetterich equation without extra terms because the background field
$\phi^0$ is scale-independent. 
For finite $\Lambda$, the vertex expansion of  $\tilde{\Gamma}_{\Lambda} [ \varphi ]$
is then of the form
 \begin{eqnarray}
 \tilde{\Gamma}_{\Lambda} [ \varphi ] & = & \tilde{\Gamma}^{(0)}_{\Lambda} +
 \tilde{\Gamma}^{(1)}_{\Lambda} \varphi_{Q=0} 
 + \frac{1}{2} \int_Q \tilde{\Gamma}^{(2)}_{\Lambda} ( Q ) \varphi_{-Q} \varphi_Q
 \nonumber
 \\
 & + &
 \sum_{ n =3}^{\infty} \frac{1}{n!} \int_{Q_1} \cdots \int_{Q_n}
 \delta ( Q_1 + \ldots + Q_n ) 
 \nonumber
 \\
 & & \hspace{10mm} \times \tilde{\Gamma}_\Lambda^{(n)} ( Q_1 , \ldots , Q_n )
 \varphi_{Q_1} \cdots \varphi_{Q_n}.
 \hspace{7mm}
 \label{eq:tildeGammaexpansion}
 \end{eqnarray}
By construction, at the initial scale 
 \begin{equation}
\tilde{\Gamma}_{\Lambda_0} [ \varphi ] =  S_{\rm eff} [ \phi^0 + \varphi] ,
 \end{equation}
where the
proper initial value should be determined by the
boundary condition
 \begin{equation}
 \lim_{\Lambda \rightarrow 0} \tilde{\Gamma}^{(1)}_{\Lambda} =0.
 \label{eq:boundarycondition}
 \end{equation}
The initial vertex 
$\tilde{\Gamma}^{(1)}_{\Lambda_0}$ is then given by
 \begin{equation}
 \tilde{\Gamma}^{(1)}_{\Lambda_0} =    \omega_0^2 \phi^0 + \gamma_0 \int_K \tilde{G}_0 ( K ).
 \label{eq:onepoint1}
 \end{equation}
Here 
 \begin{equation}
\tilde{G}_0 ( K ) =
 \frac{1}{i \omega - \epsilon_{\bd{k}} + \mu -   \gamma_0 \phi^0 }
 \end{equation}
is the fermion propagator with shifted chemical potential, where the shift
$- \gamma_0 \phi^0=   \gamma_0^2 \rho / \omega_0^2$
 takes all 
tadpole contributions to the electronic self-energy into account.
The initial conditions for all  other shifted vertices 
$\tilde{\Gamma}^{(n)}_{\Lambda_0} ( Q_1 , \ldots , Q_n )$ with $ n \geq 2$ can be obtained from the
corresponding unshifted expressions 
$\Gamma^{(n) }_{\Lambda_0} ( Q_1 , \ldots , Q_n )$ by replacing $G_0 ( K ) \rightarrow 
\tilde{G}_0 ( K )$ in all fermion loops. In particular,
the two-point vertex is initially given by
 \begin{equation}
 \tilde{\Gamma}_{\Lambda_0}^{(2)} ( Q ) = \bar{\omega}^2 + \omega_0^2 +\gamma_0^2 \int_K \tilde{G}_0 ( K ) 
 \tilde{G}_0 ( K + Q ).
 \end{equation}

\begin{figure}
\begin{center}
  \centering
\vspace{7mm}
 \includegraphics[width=0.45\textwidth]{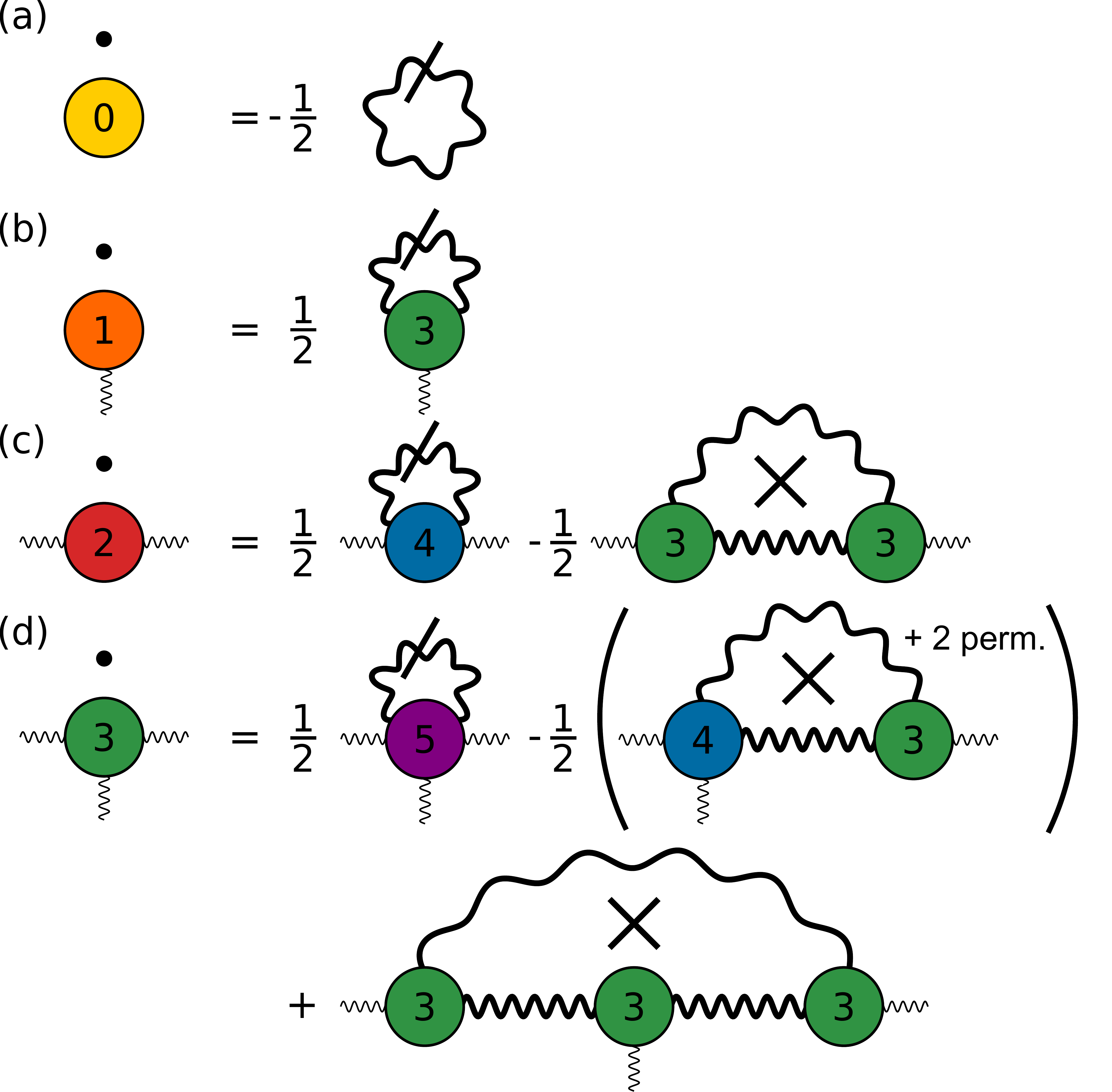}%\nspace
   \end{center}
%  \vspace{-4mm}
  \caption{%
%(Color online)
Graphical representation of exact FRG flow equations for the irreducible vertices of effective phonon action
(\ref{eq:Seffphonon}).
(a) Flow equation~(\ref{eq:zeropointflow}) for the free energy; (b) 
flow equation~(\ref{eq:onepointflow}) for the one-point vertex;
(c) flow equation~(\ref{eq:twopointflow}) for the
two-point vertex; (d) flow equation~(\ref{eq:threepointflow}) for the three-point vertex, where 
``+ 2 perm.'' denotes  two additional diagrams obtained by exchanging the labels
 $( Q_1 \leftrightarrow Q_2)$ and $( Q_1 \leftrightarrow Q_3)$ of the external legs 
attached to the three-point vertex.
The thick
wavy lines represent the regularized phonon propagator $\tilde{D}_\Lambda ( Q )$, while the slash represents the regulator insertion, $-\partial_{\Lambda} R_{\Lambda} ( Q)$. A cross inside a loop corresponds to a sum where 
each propagator of the loop is once slashed according to the product rule (\ref{eq:productrule}).
 A colored circle labeled by the number $n$ represents an $n$-point vertex $\tilde{\Gamma}^{(n)}_\Lambda$. The dots above the vertices denote a scale derivative.} 
\label{fig:flowequations}
\end{figure}

To obtain the exact flow equation for the free energy (in units of temperature)
we set $\phi = \phi^0$ in the Wetterich equation (\ref {eq:Wetterich1}) 
and obtain
 \begin{equation}
 \partial_{\Lambda} \tilde{\Gamma}^{(0)}_{\Lambda} = \frac{1}{2} \int_Q \tilde{D}_{\Lambda} ( Q )
 \partial_{\Lambda} R_{\Lambda} ( Q ),
 \label{eq:zeropointflow}
 \end{equation}
where
 \begin{equation}
 \tilde{D}_{\Lambda} ( Q ) = \frac{1}{ \tilde{\Gamma}^{(2)}_\Lambda ( Q ) + R_{\Lambda} ( Q ) }
 \label{eq:phonreg}
 \end{equation}
is the regularized phonon propagator.
Next, we substitute the expansion (\ref{eq:tildeGammaexpansion}) 
into Eq.~(\ref{eq:Wetterich1}) and compare the coefficients on both sides to 
obtain
 \begin{equation}
  \partial_{\Lambda} \tilde{\Gamma}^{(1)}_{\Lambda}  =
   \frac{1}{2} \int_Q \dot{\tilde{D}}_{\Lambda} ( Q ) \tilde{\Gamma}^{(3)}_{\Lambda} ( - Q , Q , 0),
 \label{eq:onepointflow}
 \end{equation}
where the single-scale propagator
is defined by
 \begin{equation}
 \dot{ \tilde{D}}_{\Lambda} ( Q ) = - \tilde{D}_{\Lambda}^2 ( Q ) \partial_{\Lambda} R_{\Lambda} ( Q).
 \end{equation}
Similarly, we find that  the two-point vertex satisfies the exact flow equation
 \begin{eqnarray}
 & & \partial_{\Lambda} \tilde{\Gamma}^{(2)}_{\Lambda} ( Q )  =  
%\tilde{\Gamma}_{\Lambda}^{(3)}
% (-Q , Q , 0) \partial_{\Lambda} X^0_{\Lambda}
% \nonumber
% \\
 \frac{1}{2} \int_{ Q^{\prime} } \dot{\tilde{D}}_{\Lambda} ( Q^{\prime} ) 
 \tilde{\Gamma}^{(4)}_{\Lambda} ( - Q^{\prime} , Q^{\prime} , -Q , Q )
 \nonumber
 \\
 &  & - \frac{1}{2} \int_{ Q^{\prime}} [  {\tilde{D}}_{\Lambda} ( Q^{\prime} ) 
 \tilde{D}_{\Lambda} ( Q^{\prime} + Q )]^{\bullet}
  \tilde{\Gamma}^{(3)}_{\Lambda} ( - Q ,   Q+ Q^{\prime} ,-  Q^{\prime} )
 \nonumber
 \\
 & & \hspace{7mm} \times 
\tilde{\Gamma}^{(3)}_{\Lambda} ( Q^{\prime} ,  -  Q^{\prime}  - Q , Q ),
 \label{eq:twopointflow}\\
\nonumber
 \end{eqnarray}
where we have introduced the product notation for single-scale propagators,
 \begin{eqnarray}
 & & [  {\tilde{D}}_{\Lambda} ( Q^{\prime} ) \tilde{D}_{\Lambda}
 (  Q^{\prime} + Q )]^{\bullet}
 \nonumber
 \\
 & = &   \dot{\tilde{D}}_{\Lambda} ( Q^{\prime} ) \tilde{D}_{\Lambda}
 ( Q^{\prime} + Q ) +  {\tilde{D}}_{\Lambda} ( Q^{\prime} ) \dot{\tilde{D}}_{\Lambda}
 (  Q^{\prime} + Q ). \hspace{7mm}
 \label{eq:productrule}
 \end{eqnarray}
Finally, we also need the flow of the three-point vertex which is given by
 \begin{widetext}
 \begin{eqnarray}
& &  \partial_{\Lambda} \tilde{\Gamma}^{(3)}_{\Lambda} ( Q_1 , Q_2 , Q_3 )
  =  
%\tilde{\Gamma}_{\Lambda}^{(4)}
% (Q_1 , Q_2 , Q_3, 0) \partial_{\Lambda} X^0_{\Lambda} 
  \frac{1}{2} \int_{ Q} \dot{\tilde{D}}_{\Lambda} ( Q ) 
 \tilde{\Gamma}^{(5)}_{\Lambda} ( - Q , Q , Q_1, Q_2, Q_3 )
 \nonumber
 \\
 &  & -  \frac{1}{2} \int_Q     \left\{   [  {\tilde{D}}_{\Lambda} ( Q ) 
 \tilde{D}_{\Lambda} ( Q + Q_1 )]^{\bullet}  
 \tilde{\Gamma}^{(3)}_{\Lambda} ( Q_1, Q  , - Q_1 - Q )
  \tilde{\Gamma}^{(4)}_{\Lambda} ( -  Q  , Q_1 + Q , Q_2 , Q_3)
+ ( Q_1 \leftrightarrow Q_2) + ( Q_1 \leftrightarrow Q_3) 
 \right\}
 \nonumber
 \\
 & & + \int_Q   [  {\tilde{D}}_{\Lambda} ( Q ) 
 \tilde{D}_{\Lambda} ( Q + Q_1 )  \tilde{D}_{\Lambda} ( Q - Q_2 )  ]^{\bullet}  
 \tilde{\Gamma}^{(3)}_{\Lambda} ( Q_1, Q, - Q_1 - Q )
 \tilde{\Gamma}^{(3)}_{\Lambda} ( Q_2, - Q, - Q_1 + Q )
\tilde{\Gamma}^{(3)}_{\Lambda} ( Q_3,  Q_1 + Q , -Q + Q_2 ).
 \nonumber
 \\
 & &
 \label{eq:threepointflow}
\end{eqnarray}
 \end{widetext}
Diagrammatic representations of the exact FRG flow equations 
(\ref{eq:zeropointflow}), (\ref{eq:onepointflow}), (\ref{eq:twopointflow}), and
(\ref{eq:threepointflow})
are shown in Fig.~\ref{fig:flowequations}.

\subsection{Classification of couplings and
flow of relevant couplings}
 \label{sec:class}

To classify the infinite set of vertices in our scale-dependent 
average effective
phonon action $\tilde{\Gamma}_{\Lambda} [ \varphi ]$, we note
that 
for small momenta $ q \ll k_F$ and for frequencies $ | \bar{\omega} | \ll v_F q $
the two-point vertex is initially given by
  \begin{equation}
 \tilde{\Gamma}^{(2)}_{\Lambda_0} ( Q ) = r_{0} + b_{0} | \bar{\omega} | / q 
 + c_{0} q^2 + {\cal{O}} ( \bar{\omega}^2  , q^4 ),
 \label{eq:Gamma2init}
 \end{equation}
where the  $r_0$, $b_0$ and $c_0$
are the RPA coefficients given in Eqs.~(\ref{eq:rbcdef}).
Assuming that this form is not changed by the induced interactions between the 
phonons, the flowing 
two-point vertex at scale $\Lambda$ is given by~\cite{footnote1}
 \begin{equation}
 \tilde{\Gamma}^{(2)}_{\Lambda} ( Q ) = r_{\Lambda} + b_{\Lambda} | \bar{\omega} | / q 
 + c_{\Lambda} q^2 + {\cal{O}} ( \bar{\omega}^2  , q^4 ),
 \label{eq:Gamma2low}
 \end{equation}
where $r_{\Lambda}$ can be identified with  the square of the renormalized phonon frequency at scale $\Lambda$. Note that the correction of order $\bar{\omega}^2$ 
due to the inverse bare propagator
is negligible relative to the Landau damping term
$b_0 |  \bar{\omega} | / q $ if
 \begin{equation}
 q \lesssim  \sqrt{\lambda_0}  \frac{ {\omega}_0 }{ v_F} = 
 \frac{ \gamma_0 \sqrt{ \nu} }{v_F} \equiv q_0,
 \label{eq:q0def} 
\end{equation}
which plays the role of an 
ultraviolet cutoff in  our low-energy theory.
In the adiabatic limit $ \omega_0 \ll \epsilon_F$ the cutoff
$q_0$ is small compared with $k_F$,  
while $q_0 \gg k_F$ in the antiadiabatic limit $\omega_0 \gg \epsilon_F$. 

If the system exhibits a Pomeranchuk instability then the coupling 
$r_{\Lambda}$ vanishes for $\Lambda \rightarrow 0$. 
It is then natural to rescale momenta, frequencies, and the field such that 
the couplings $b_{\Lambda} $ and $c_{\Lambda}$ are both marginal, which is achieved by rescaling them
by the following powers of $\Lambda$,
 \begin{subequations}
 \begin{eqnarray}
 q & \propto & \Lambda,
 \\
 \bar{\omega} & \propto & \Lambda^z,
 \\
 \phi_Q & \propto &  \Lambda^{-1 - \frac{ d+z}{2}},
 \end{eqnarray}
 \end{subequations}
where we introduce the dynamical exponent
 \begin{equation}
 z =3.
 \end{equation}
Note that the powers of $\Lambda$ are simply the canonical dimensions of the
corresponding quantities, which can be determined by
dimensional analysis. 
The one-point vertex inherits the canonical scaling of the field, 
 \begin{equation}
 \tilde{\Gamma}^{(1)}_{\Lambda} \propto  \Lambda^{-1 - \frac{ d+z}{2}}
 = \Lambda^{ - \frac{d+5}{2} },
 \end{equation}
while  the three-point and the four-point vertices for vanishing energy-momenta scale as
follows,
 \begin{subequations}
 \begin{eqnarray}
 \tilde{\Gamma}^{(3)}_{\Lambda} (0,0,0 ) & \propto &  \Lambda^{ -3 + \frac{ d+z}{2}} =
 \Lambda^{ \frac{d-3}{2}}
,
 \\
 \tilde{\Gamma}^{(4)}_{\Lambda} (0,0,0,0 ) & \propto &  \Lambda^{ -4 +  d+z} = \Lambda^{d-1}.
 \end{eqnarray}
 \end{subequations}
The important point is that the three-point vertex is relevant below three dimensions and 
therefore cannot be neglected in this case. 
Keeping in mind that with the dynamic exponent 
$z=3$ the effective dimensionality is  shifted to  $d + z = d+3$ \cite{Hertz76},
the relevance of the three-point vertex in our effective phonon theory is consistent with the well-known
fact that in $\phi^3$-theory
the three-point vertex becomes relevant below
six dimensions \cite{Polyakov70,Migdal71,Mack73,Fisher78,Fei14,Rong20}. 
To take into account the most relevant interaction processes
between the phonons in dimensions $d \leq 3$  
we therefore should consider the projected RG flow in the space of the following three couplings,
 \begin{subequations}
 \begin{eqnarray}
  h_{\Lambda} & = & \tilde{\Gamma}^{(1)}_{\Lambda} ,
 \\
 r_{\Lambda} & = & \tilde{\Gamma}^{(2)}_{\Lambda} ( 0 ),
 \\
 g_{\Lambda} & = & \tilde{\Gamma}^{(3)}_{\Lambda} ( 0,0,0).
 \end{eqnarray}
 \end{subequations}
Neglecting all other couplings,
the exact flow equations (\ref{eq:onepointflow}), (\ref{eq:twopointflow}), and
(\ref{eq:threepointflow}) reduce to the following system of truncated flow equations for the relevant couplings,
 \begin{subequations}
 \label{eq:flowhrg}
 \begin{eqnarray}
 \partial_{\Lambda} h_{\Lambda} & = &   \frac{ {g}_{\Lambda}}{2}
 \int_Q \dot{\tilde{D}}_{\Lambda} ( Q ),
 \label{eq:flowh}
 \\
 \partial_{\Lambda} {r}_{\Lambda} & = & - {g}_{\Lambda}^2 
 \int_Q \dot{\tilde{D}}_{\Lambda} ( Q ) \tilde{D}_{\Lambda} ( Q ),
 \label{eq:flowtruncr}
 \\
 \partial_{\Lambda} {g}_{\Lambda} & = & 3  {g}_{\Lambda}^3 
 \int_Q \dot{\tilde{D}}_{\Lambda} ( Q ) \tilde{D}^2_{\Lambda} ( Q ).
 \label{eq:flowg}
 \end{eqnarray}
 \end{subequations}
In the following, we will neglect the flow of the marginal couplings
$b_{\Lambda}$ and $c_{\Lambda}$ in the low-energy expansion (\ref{eq:Gamma2low}),
which amounts to neglecting the 
momentum and frequency dependence of the phonon self-energy,
$\Delta_{\Lambda} ( Q ) \approx \Delta_{\Lambda} (0)$. We thus  approximate
$b_{\Lambda} \approx b_0$ and $c_{\Lambda} \approx c_0$, where the
initial values $b_0$ and $c_0$ are given in Eqs.~(\ref{eq:b0def}) and
(\ref{eq:c0def}).
Within this truncation the possibility that the dynamical exponent $z$
is modified by an anomalous dimension 
is not taken into account.
The regularized flowing phonon propagator is then given by
 \begin{equation}
 \tilde{D}_{\Lambda} ( Q ) = \frac{1}{r_{\Lambda} + b_0 | \bar{\omega} | / q 
 + c_0 q^2 + R_{\Lambda} ( Q ) }.
 \end{equation}
At this point we have to specify the regulator. While the required boundary conditions
can be satisfied in many ways,
for our purpose it is most convenient to work with a Litim-type regulator \cite{Litim01}
adapted to the peculiar momentum and frequency dependence of the bare propagator,
 \begin{equation}
 R_{\Lambda} ( Q ) = ( c_0 \Lambda^2   - b_0  | \bar{\omega} | / q  - c_0 q^2  )
 \Theta ( c_0  \Lambda^2   -  b_0  | \bar{\omega}  | / q  - c_0 q^2  ).
 \end{equation}
The momentum and frequency integrations in our flow equations (\ref{eq:flowhrg})
can then be carried out exactly and we obtain
\begin{subequations}
 \label{eq:flowhrgLitim}
 \begin{eqnarray}
 \partial_\Lambda h_\Lambda & = & -   \frac{2 K_d}{\pi ( d+1)(d+3)}
 \frac{c_0^2}{b_0} 
 \frac{ \Lambda^{d+4} g_{\Lambda}}{   ( r_{\Lambda} + c_0 \Lambda^2)^2 },
 \\
 \partial_{\Lambda} r_{\Lambda} & = & \frac{4 K_d}{\pi ( d+1)(d+3)} \frac{c^2_0}{b_0} 
 \frac{ \Lambda^{d+4} g^2_{\Lambda}}{  ( r_{\Lambda} + c_0 \Lambda^2 )^3},
 \\
 \partial_{\Lambda} g_{\Lambda} & = & - \frac{12 K_d}{\pi (d+1 )(d+3)} \frac{c^2_0}{b_0} 
 \frac{ \Lambda^{d+4}   g_{\Lambda}^3}{ ( r_\Lambda + c_0 \Lambda^2 )^4},
 \hspace{7mm}
 \end{eqnarray}
 \end{subequations}
where $K_d$ is the surface area of the $d$-dimensional unit sphere divided by $(2 \pi)^d$.
Introducing the logarithmic flow parameter $l = \ln ( \Lambda_0 / \Lambda )$ and
the dimensionless rescaled couplings
\begin{subequations}
 \begin{eqnarray}
 \tilde{h}_l & = & \frac{h_{\Lambda}}{ \sqrt{  \frac{ K_d }{\pi ( d+1)(d+3)} } \frac{c_0}{\sqrt{b_0}}  
 \Lambda^{ \frac{d+5}{2}} },
 \\
 \tilde{r}_l & = & \frac{r_{\Lambda}}{ c_0 \Lambda^2},
 \\
 \tilde{g}_l & = & \sqrt{ \frac{4 K_d}{\pi (d+1)(d+3)}} \frac{ \Lambda^{\frac{d-3}{2}} g_{\Lambda}}{ 
 c_0 \sqrt{b_0} },
 \label{eq:tildegl}
 \end{eqnarray}
 \end{subequations}
the  flow equations (\ref{eq:flowhrgLitim})  can be written as
 \begin{subequations}
 \label{eq:flowsys}
 \begin{eqnarray}
 \partial_l \tilde{h}_l & = & \frac{ d+5}{2} \tilde{h}_l +   \frac{\tilde{g}_l}{  ( \tilde{r}_l +1)^2 },
 \label{eq:flowhl} 
\\
  \partial_l \tilde{r}_l & = & 2 \tilde{r}_l - \frac{\tilde{g}_l^2}{  (\tilde{r}_l +1 )^3} ,
 \label{eq:flowr}
 \\
\partial_l \tilde{g}_l  & = &   \frac{ 3 -d }{2} \tilde{g}_l +  \frac{   3 \tilde{g}_l^3}{ ( \tilde{r}_l +1 )^4 }.
 \label{eq:flowgl}
 \end{eqnarray}
 \end{subequations}
Using the RPA expressions~(\ref{eq:rbcdef}) and the initial condition~(\ref{eq:onepoint1})
we find that the initial values of our rescaled couplings are
 \begin{subequations}
 \begin{eqnarray}
 \tilde{h}_0 & = & \frac{ 
 \frac{\rho_0 ( \mu + \delta \mu )}{\nu \epsilon_F }
 - \frac{\delta \mu }{ \lambda_0 \epsilon_F }}{ 
 \sqrt{ \frac{K_d C_d^2}{ \pi (d+1) (d+3) B_d}}     
 \sqrt{ \frac{ 2 k_F^d}{ \nu \epsilon_F}}
 \left( \frac{\Lambda_0 }{ k_F } \right)^{ \frac{ d+5}{2}} },
\\
 \tilde{r}_0 & = & 
 \frac{1- \lambda_0}{C_d \lambda_0}
 \left( \frac{k_F}{\Lambda_0} \right)^2,
 \\
 \tilde{g}_0
 & = &  \sqrt{ \frac{4 K_d}{ \pi (d+1) (d+3) B_d C_d^2} }
 \sqrt{  \frac{2 k_F^d}{\nu \epsilon_F} }  \frac{ \nu^{\prime} \epsilon_F}{\nu}
  \left( \frac{ \Lambda_0 }{k_F } \right)^{\frac{d-3}{2}}.\nonumber\\
 \hspace{7mm}
 \end{eqnarray}
 \end{subequations}
Here $\delta \mu = - \gamma_0 \phi^0$ is the shift of the chemical potential due to the
finite expectation value of the phonon displacement and
$\rho_0 ( \mu + \delta \mu ) = \int_K \tilde{G}_0 ( K )$ is the density of free electrons at the
shifted chemical potential $\mu + \delta \mu$. Note that $\delta \mu$ should be considered as a free parameter which should be adjusted such that  the coupling $h_{\Lambda}$ 
vanishes for $\Lambda \rightarrow 0$, which enforces the boundary condition \eqref{eq:boundarycondition}.

\subsection{RG flow and Pomeranchuk fixed points in $d > 3$}
\label{sec:FRGphononRGflow}

\begin{figure*}
% \begin{center}
%  \centering
\vspace{7mm}
 \includegraphics[width=0.32\textwidth]{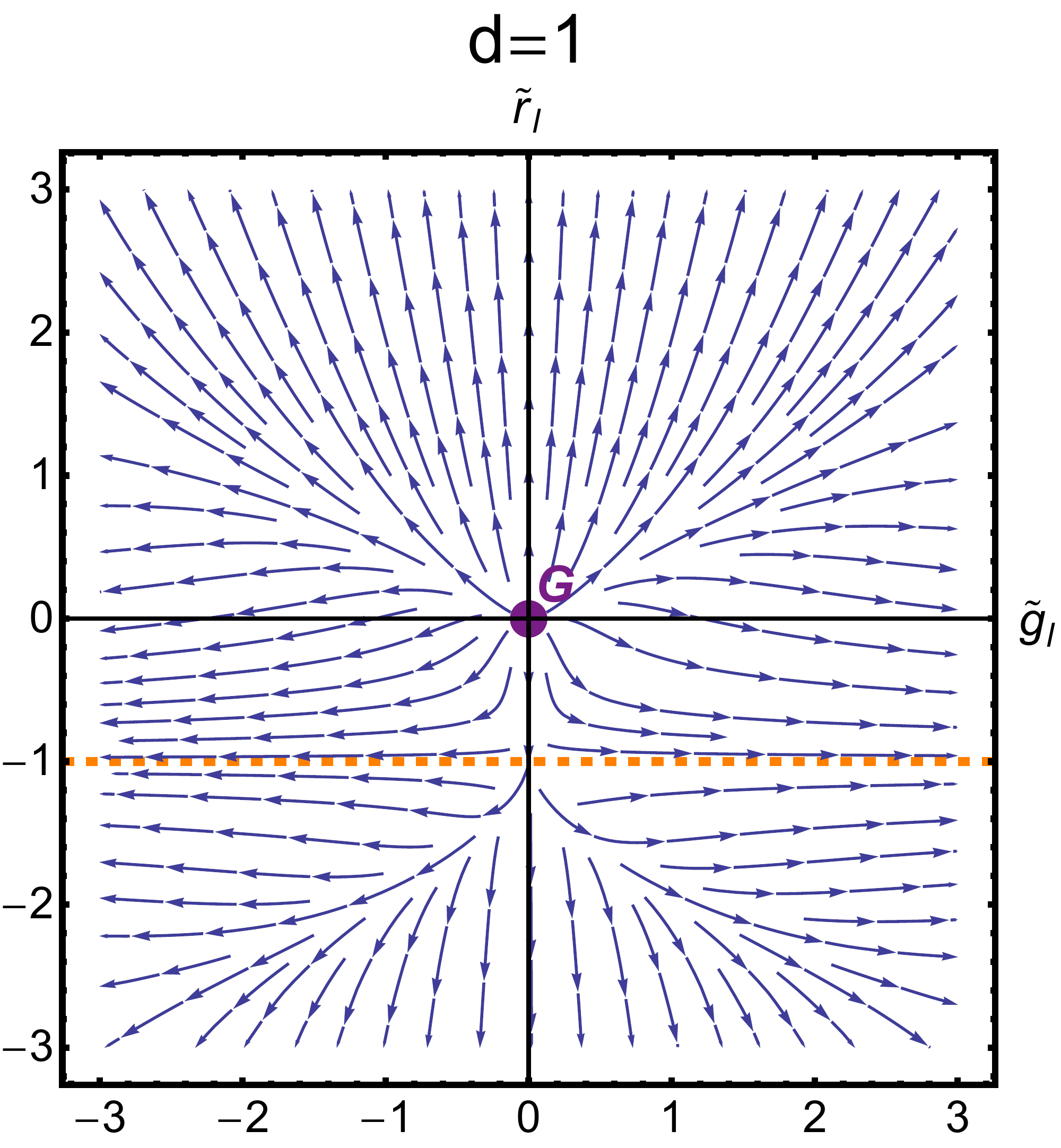}%\nspace
\includegraphics[width=0.32\textwidth]{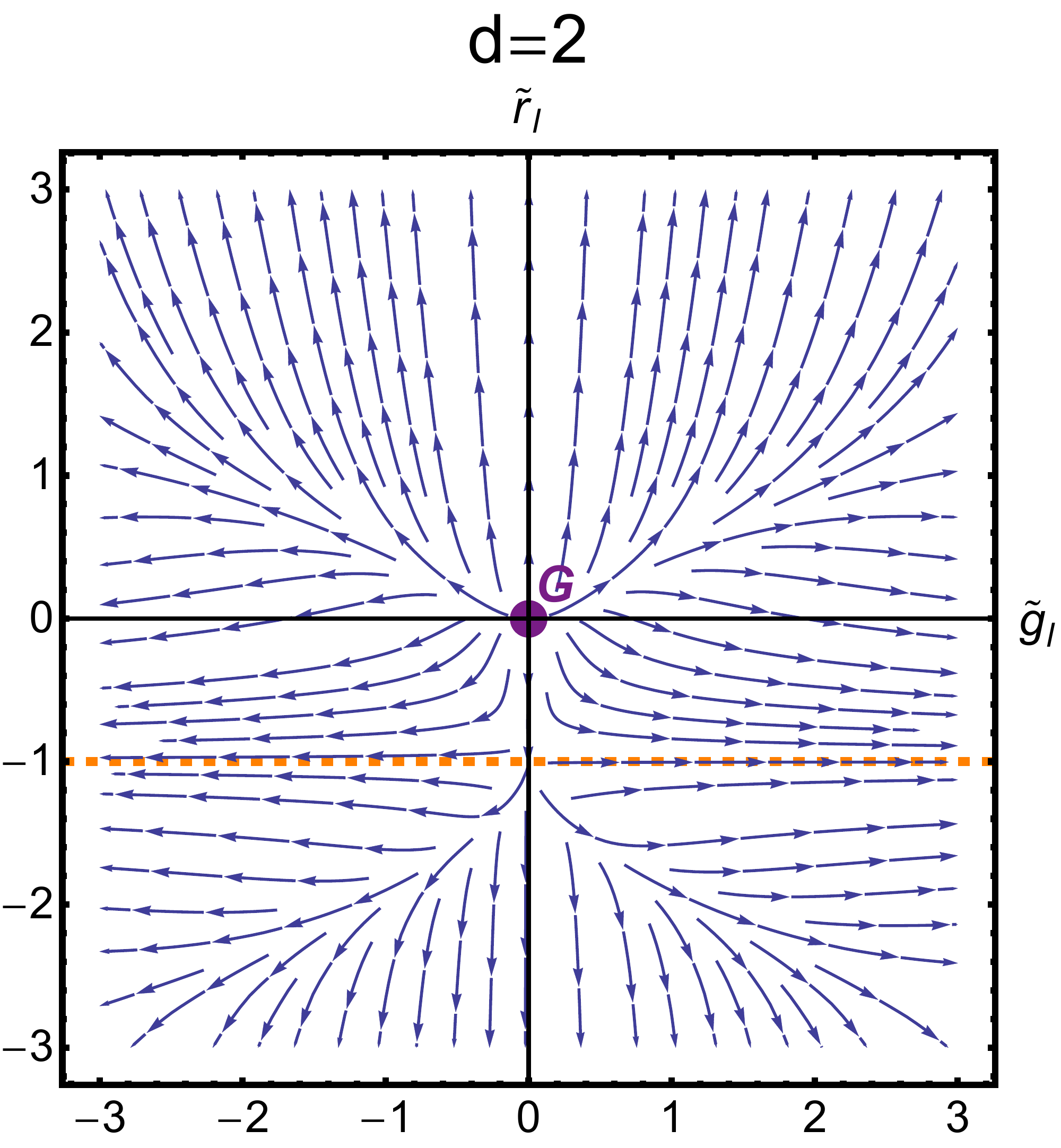}%\nspace
\includegraphics[width=0.32\textwidth]{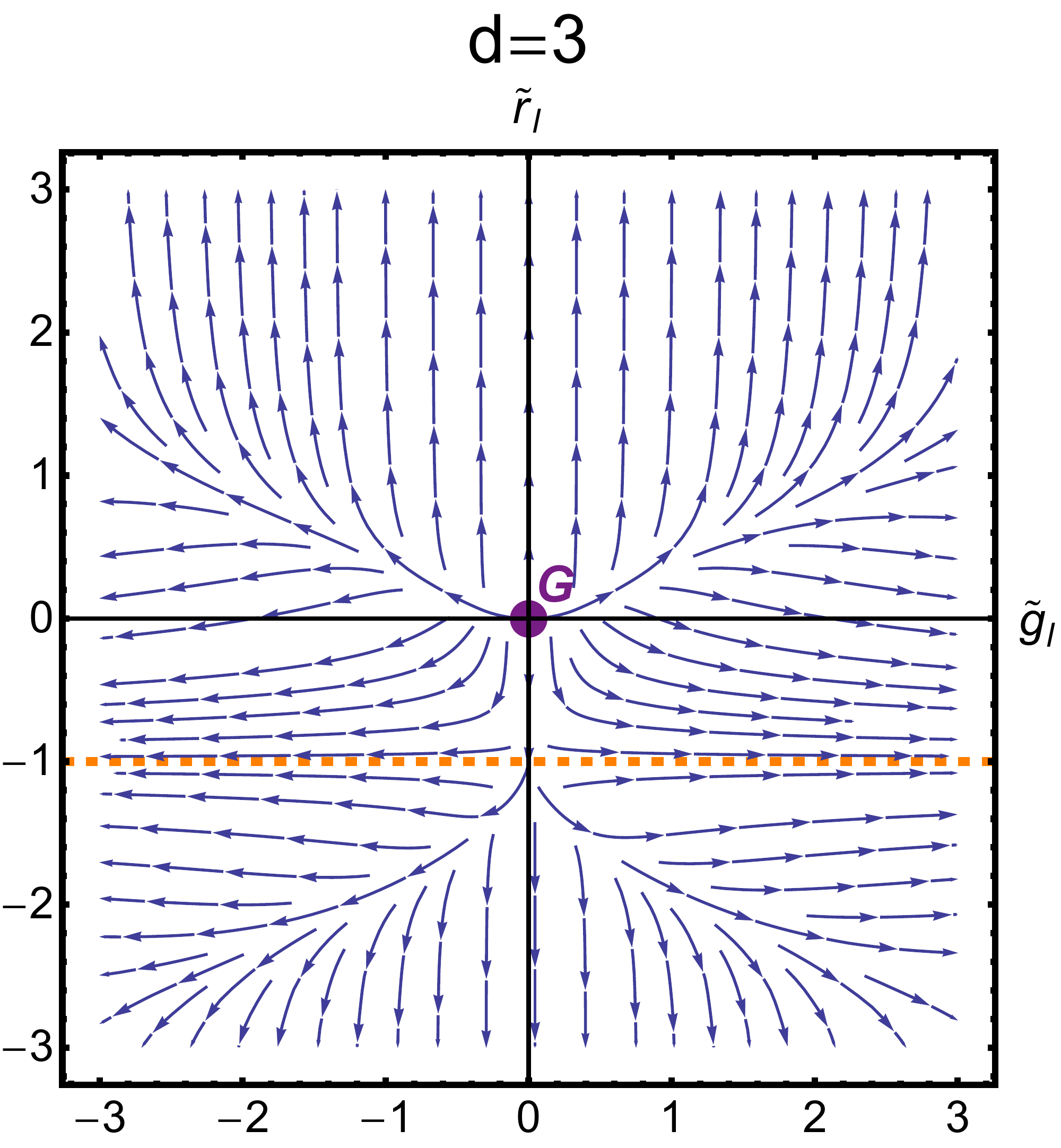}%\nspace  
%  \vspace{-4mm}
  \caption{%
%(Color online)
RG flow of the effective phonon action (\ref{eq:Seffphonon}) in the $\tilde{g}_l - \tilde{r}_l$-plane in $d=1,2,3$ obtained from the numerical solution of the flow equations (\ref{eq:flowsys}). 
The arrows indicate the direction of the RG flow toward the infrared.
For $d\leq 3$ the dimensionless rescaled flow equations have only a trivial, i.e., Gaussian, fixed point $G$, which is represented by a purple dot.
A subset of initial values, which exhibit a runaway flow, tend to $\tilde{r}_{\ast}=-1$ which is marked as an orange dashed line. }
\label{fig:RGflow123}
% \end{center}
\end{figure*}

In Fig.~\ref{fig:RGflow123} we show 
the RG flow of our effective phonon action obtained from the numerical solution
of the flow equations (\ref{eq:flowsys}) for $d =1,2,3$.
%
%\begin{widetext}
% \end{widetext}
%
Obviously, the RG flow has only a trivial Gaussian fixed point $G$
where all couplings vanish;  nontrivial fixed points with finite values of rescaled couplings
do not exist in $d \leq 3$.
We conclude that in and below three dimensions the Holstein model does not have a quantum critical point associated with a Pomeranchuk instability.
Note that in a perturbative calculation
the relevant three-point vertex which is essential for this result appears only at order $\lambda_0^3$ 
so that our calculation to order $\lambda_0^2$ in Sec.~\ref{sec:pert} does not include the relevant critical fluctuations.

\begin{figure}
 \begin{center}
  \centering
%\vspace{7mm}
% \includegraphics[width=0.9\textwidth]{fig_Litimflow.pdf}%\nspace
%\includegraphics[width=0.9\textwidth]{fig_Litimflowepsilon.pdf}%\nspace
%\includegraphics[width=0.5\textwidth]{fig_Litimflowepsilon.pdf}%\nspace
% \\
%
\includegraphics[width=0.32\textwidth]{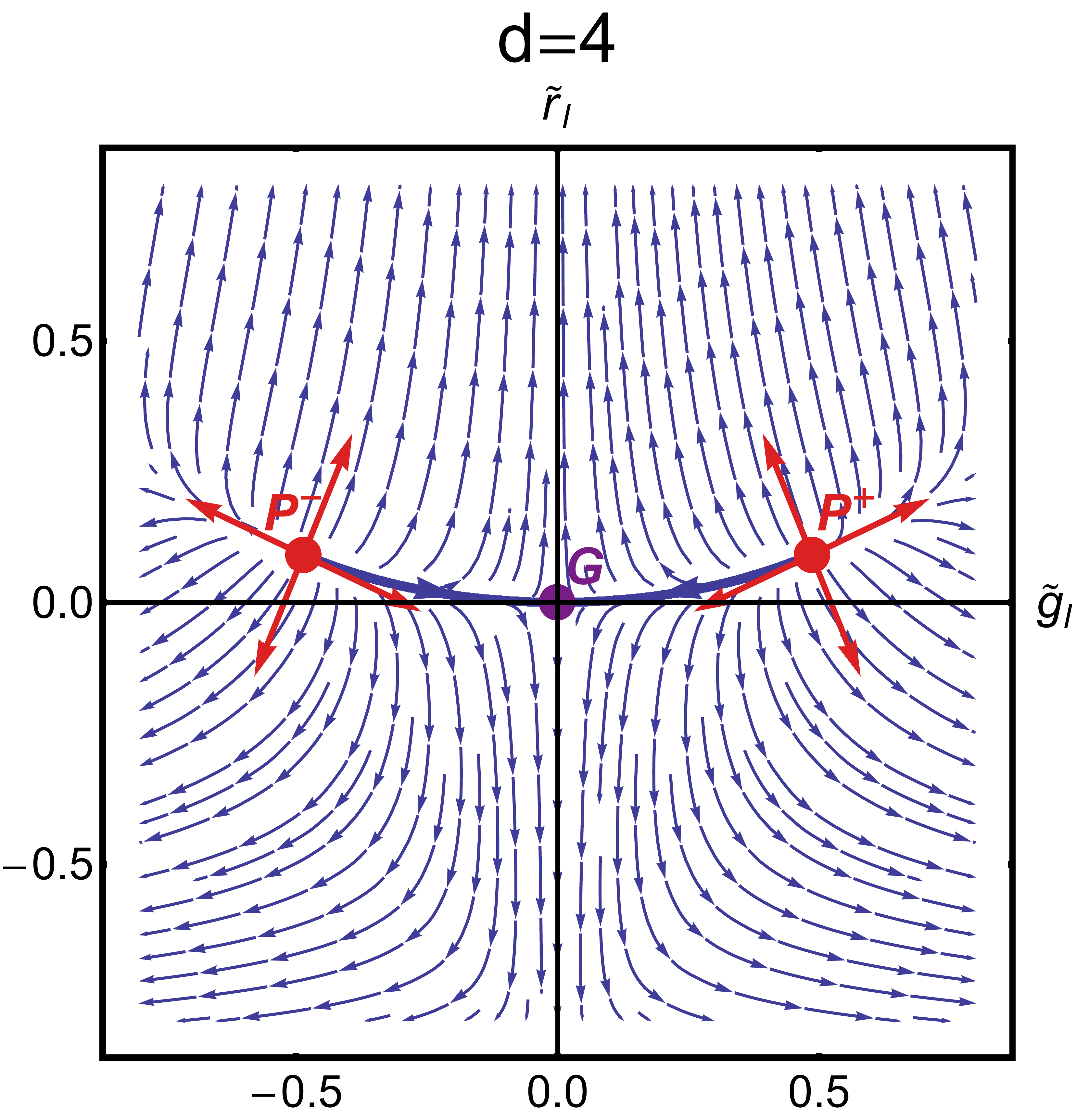}%\nspace
   \end{center}
%  \vspace{-4mm}
  \caption{%
%(Color online)
RG flow of the effective phonon action (\ref{eq:Seffphonon}) in the $\tilde{g}_l-\tilde{r}_l$ plane in $d=4$ obtained from the numerical solution of the flow equations (\ref{eq:flowsys}). 
In addition to the Gaussian fixed point $G$ there are two nontrivial fixed points $P^-$ and $P^+$, which are denoted by red dots, corresponding to a Pomeranchuk instability. The two Pomeranchuk fixed points are tricritical, as there are three relevant couplings $\tilde{h}_l$,
$\tilde{r}_l$, and $\tilde{g}_l$  whose initial values have to be fine-tuned to reach the fixed point.
The eigenvectors of the linearized flow around the Pomeranchuk fixed point are represented by red arrows. Note only the eigenvectors which lie in the $\tilde{g}_l$-$\tilde{r}_l$-plane are shown. The thick blue line emanates from Pomeranchuck fixed point and flows into the Gaussian fixed point.}

\label{fig:RGflowd4}
\end{figure}

However, for $d > 3$ we obtain two nontrivial
fixed points $P^+$ and $P^-$ which can be  associated with a quantum critical point
due to  a Pomeranchuk instability. To calculate the couplings 
$ ( \tilde{h}_{\ast} , \tilde{r}_{\ast} , \tilde{g}_{\ast})$ at the fixed points, 
we set the left-hand sides of the flow equations (\ref{eq:flowsys}) 
equal to zero. The fixed-point condition due to the flow equation~(\ref{eq:flowg}) for the three-point vertex reads
 \begin{equation}
 \tilde{g}_{\ast}^2 = \frac{ d-3}{6} ( \tilde{r}_{\ast} +1 )^4,
 \label{eq:fixg}
 \end{equation}
 which is valid for $\tilde{g}_{\ast} \neq 0$. This equation has real solutions only for 
$d > 3$, hence a nontrivial fixed point
corresponding to a Pomeranchuk instability can only exist in dimensions larger than three. Similarly, the fixed point condition resulting from the flow equation (\ref{eq:flowr}) gives
 \begin{equation}
 2 \tilde{r}_{\ast} = \frac{ \tilde{g}_{\ast}^2}{ ( \tilde{r}_{\ast} +1 )^3 }.
 \label{eq:fixr}
 \end{equation}
Eliminating  $\tilde{g}_{\ast}^2$ in Eq.~(\ref{eq:fixr}) using Eq.~(\ref{eq:fixg})
we obtain an equation for the value of $\tilde{r}_{\ast}$ dependent only on the deviation $\epsilon$ from three dimensions
 \begin{equation}
 \tilde{r}_{\ast} = \frac{ \epsilon}{ 12 - \epsilon }, \; \; \; \epsilon = d-3.
 \label{eq:fixrepsilon}
 \end{equation} 
Substituting this back into Eq.~(\ref{eq:fixg}) we get 
two nontrivial fixed-point values for the
three-point vertex in dimensions $d > 3$, 
 \begin{equation}
 \tilde{g}_{\ast}^{\pm} = \pm \frac{ 24 \sqrt{ 6 \epsilon}}{( \epsilon - 	12 )^{2}} .
 \label{eq:fixgepsilon}
 \end{equation}
Finally, we combine the fixed point condition resulting 
from the flow equation
(\ref{eq:flowhl}) with the previously calculated values $\tilde{r}_{\ast}$ and $\tilde{g}_{\ast}^{\pm}$ to obtain the corresponding
fixed-point values of the rescaled one-point vertex
 \begin{eqnarray}
 \tilde{h}_{\ast}^{\pm} & = & - \frac{2}{d+5}   \frac{ \tilde{g}^{\pm}_{\ast} }{( \tilde{r}_{\ast} +1)^2 } =  \mp  \sqrt{\frac{2\epsilon   }{ 3(\epsilon + 8)}}.
 \label{eq:fixhepsilon}
 \end{eqnarray}
We conclude that for $d > 3$ the RG flow of the Holstein model
exhibits two nontrivial fixed points 
$ P^+=( \tilde{h}_{\ast}^+ , \tilde{r}_{\ast} , \tilde{g}_{\ast}^+ )$ and
$ P^- = ( \tilde{h}_{\ast}^- , \tilde{r}_{\ast} , \tilde{g}_{\ast}^- )$. At these fixed points
the renormalized phonon frequency vanishes. The uniform compressibility then diverges so that the
fixed points describe a quantum critical point associated with a Pomeranchuk instability.
The RG flow in $d=4$ is shown in Fig.~\ref{fig:RGflowd4}. For smaller $d$ the Pomeranchuk
fixed points $P^-$ and $P^+$ move toward the Gaussian fixed point $G$ until they
merge with $G$  in $d=3$.
%
%

\begin{figure*}[ht]
\begin{center}
  \centering
\vspace{7mm}
\includegraphics[width=0.32\textwidth]{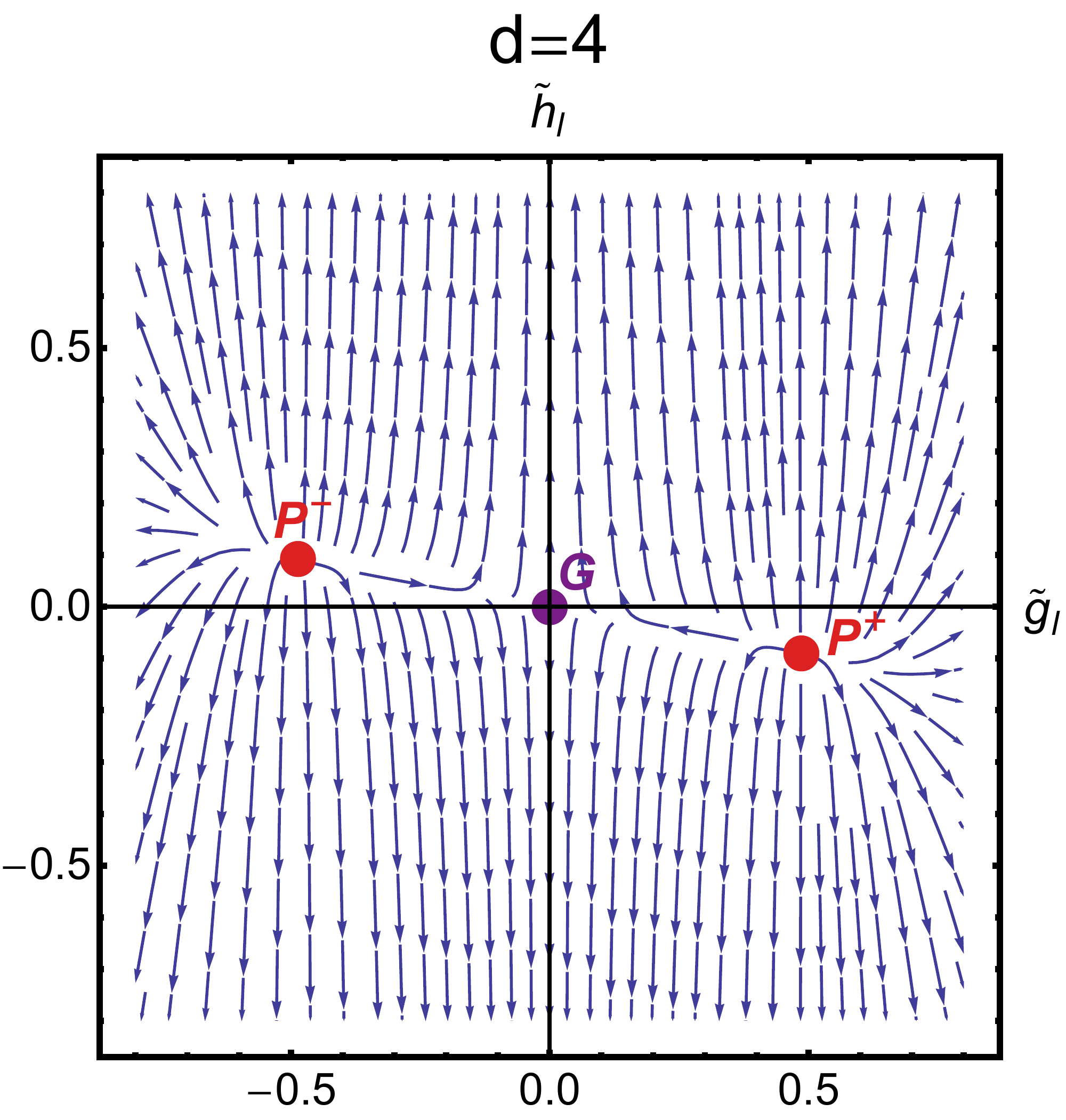}%\nspace
\includegraphics[width=0.32\textwidth]{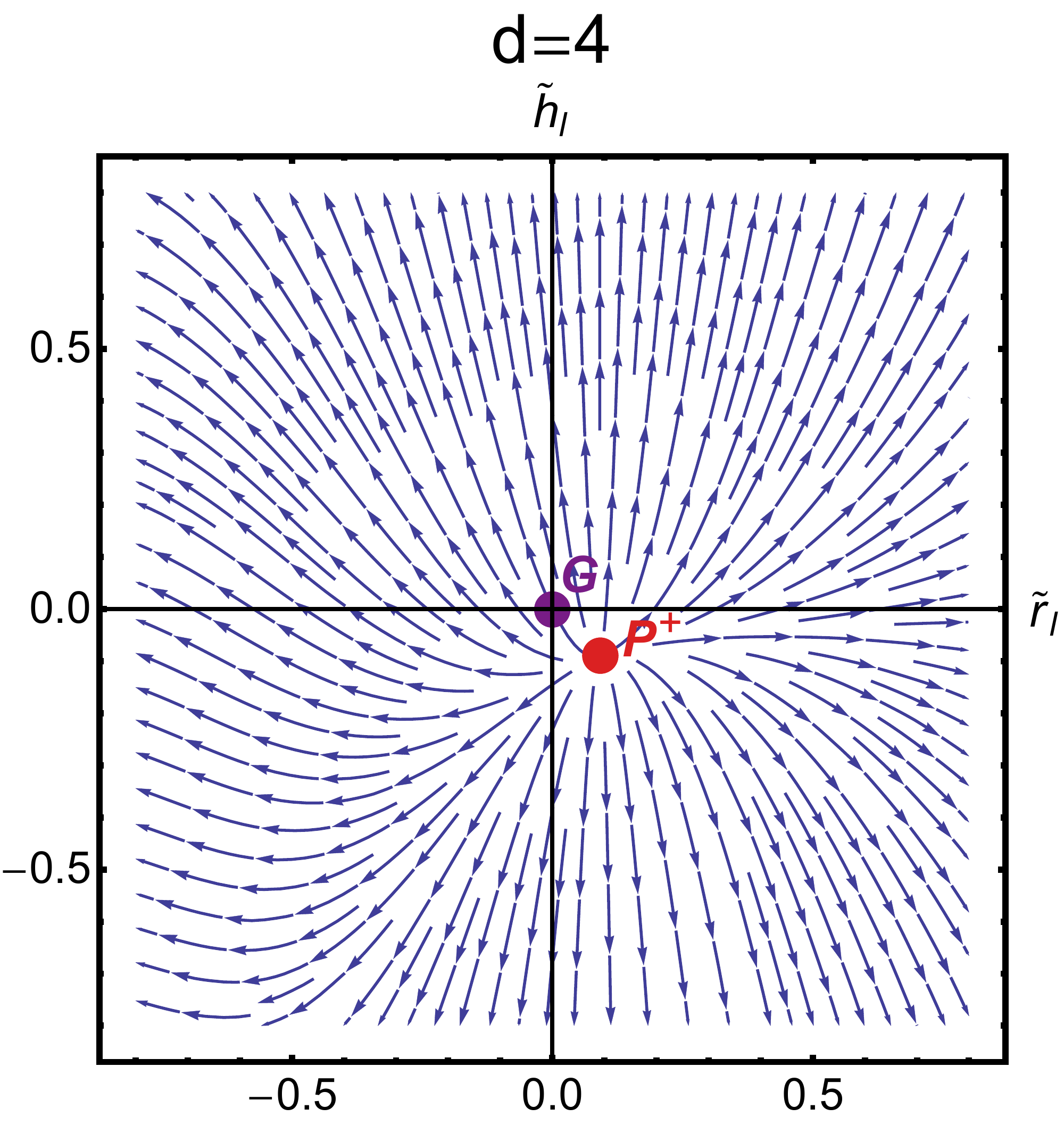}%\nspace
\includegraphics[width=0.32\textwidth]{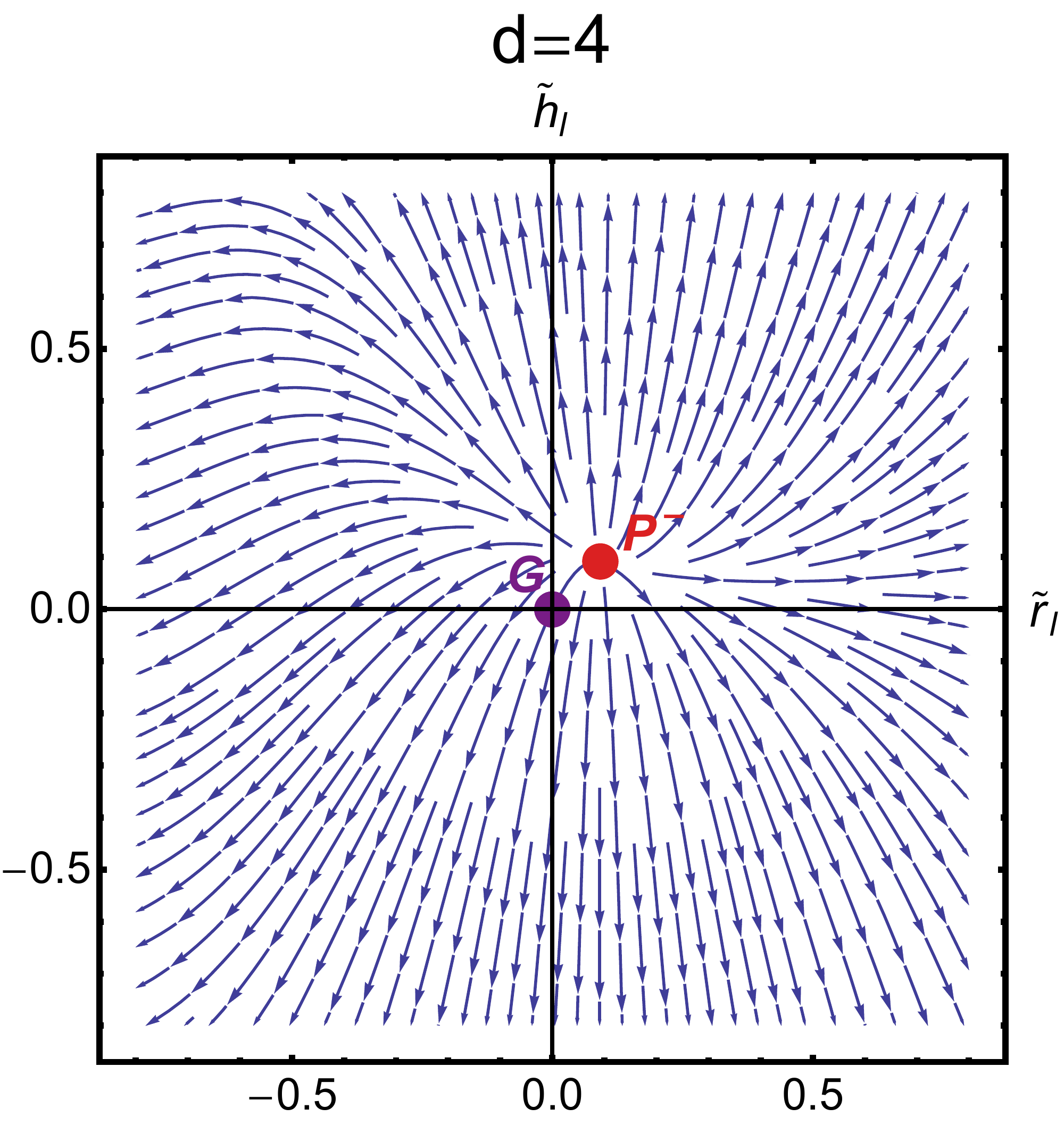}%\nspace
   \end{center}
%  \vspace{-4mm}
  \caption{%
%(Color online)
RG flow of the effective phonon action (\ref{eq:Seffphonon}) in $d=4$ obtained from the numerical solution of the flow equations (\ref{eq:flowsys}). 
%The arrows point in the direction of RG flow towards the infrared.
Left: RG flow in the $\tilde{g}_l$ - $\tilde{r}_l$ plane at $\tilde{r}=\tilde{r}_{\ast}$.
Middle:  RG flow in the $\tilde{r}_l$ - $\tilde{h}_l$ plane at $\tilde{g}_l=\tilde{g}^+_{\ast}$.
Right:  RG flow in the $\tilde{r}_l$ - $\tilde{h}_l$ plane at $\tilde{g}_l=\tilde{g}^-_{\ast}$.
}
\label{fig:RGflow4}
\end{figure*}

%
%\begin{figure}
%\begin{center}
%\centering
%\vspace{7mm}
% \includegraphics[width=0.9\textwidth]{fig_Litimflow.pdf}%\nspace
%\includegraphics[width=0.5\textwidth]{fig_Litimflowepsilon.pdf}%\nspace
%\\
%\includegraphics[width=0.5\textwidth]{fig_Litimflowd4_neu.pdf}%\nspace
%\includegraphics[width=0.9\textwidth]{fig_Litimflowd4.pdf}%\nspace
%   \end{center}
%  \vspace{-4mm}
%  \caption{%
%(Color online)
%RG flow of the effective phonon action (\ref{eq:Seffphonon}) 
%in the $\tilde{g}_l-\tilde{r}_l$-plane in $d=4$ obtained from the numerical solution of the flow equations (\ref{eq:flowsys}). The arrows point in the direction of RG flow towards the
%infrared.
%The colored line indicates the 
%location of the Pomeranchuk fixed points $P^-$ and $P^+$ as a function of $\epsilon = d-3$ for values between $\epsilon = 0\ (d=3)$ and $\epsilon=1\ (d=4)$ which is given by Eqs.~(\ref{eq:fixrepsilon}), (\ref{eq:fixgepsilon}) and (\ref{eq:fixhepsilon}). For $ d > 3$ we find three fixed points: 
%a  Gaussian fixed point $G$ and two non-trivial Pomeranchuk fixed points $P^-$ and $P^+$. 
%For $d=3$ the Pomeranchuk fixed points merge with the Gaussian fixed point which remains 
%the only  fixed point for $d \leq 3$.}
%\label{fig:RGflowepsilon}
%\end{figure}
%
%
From Fig.~\ref{fig:RGflowd4} it is obvious that
the Pomeranchuk fixed points in $d > 3$ have only relevant directions in our truncated coupling space.
 Keeping the mind that the arrows on the
flow lines indicate the flow toward the infrared  (decreasing scale $\Lambda$)
the flow toward the ultraviolet (increasing scale) is opposite to the arrows on the flow 
lines. The  Pomeranchuk fixed points are therefore  ultraviolet-stable.
To calculate the corresponding scaling variables 
we linearize the flow equations (\ref{eq:flowsys}) around the fixed points,
 \begin{subequations}
 \begin{eqnarray}
\partial_l \delta \tilde{h}_l & = & \left(4+ \frac{\epsilon}{2} \right)\delta \tilde{h}_l  -\frac{2 \tilde{g}_\ast \delta \tilde{r}_l}{(1+\tilde{r}_\ast)^3}  +  \frac{\delta \tilde{g}_l}{(1+\tilde{r}_\ast)^2},
\\
\partial_l \delta\tilde{r}_l & = & \left(2+\frac{3 \tilde{g}_*^2}{ (1 + \tilde{r}_*)^4}\right) \delta \tilde{r}_l-\frac{2 \tilde{g}_*^2}{(1+\tilde{r}_*)^3}\delta \tilde{g}_l,
\\
\partial_l \delta\tilde{g}_l & = & -\frac{12 \tilde{g}_*^3}{ (1 + \tilde{r}_*)^5} 
 \delta \tilde{r}_l-\left( \frac{\epsilon}{2}-\frac{9 \tilde{g}_*^2}{(1+ \tilde{r}_*)^4}\right)\delta \tilde{g}_l.
  \hspace{7mm}
 \end{eqnarray}
 \end{subequations}
The eigenvalues and normalized eigenvectors $\bm{v}=({v}_h , 
{v}_{r}, {v}_{g})^T$ of the linearized flow close to $P^+$ are
 \begin{subequations}
\begin{align}
\lambda_1 & =  4.5,  &\bm{v}_1^{+} = \begin{pmatrix}
-0.843 \\ 
0.448 \\ 
-0.298
\end{pmatrix}, \\ \nonumber\\[0.15cm]
\lambda_2& =  0.641,  &\bm{v}_2^{+} =\begin{pmatrix}
0 \\ 
0.432 \\ 
0.902
\end{pmatrix},\\ \nonumber\\[0.15cm]
\lambda_3& =  2.860, &\bm{v}_3^{+}=\begin{pmatrix}
0 \\ 
0.928 \\ 
-0.374
\end{pmatrix}.
\end{align}
 \end{subequations}
The linearized flow close to
$P^-$ has the same eigenvalues but different eigenvectors which can be obtained by 
inverting the first and third components of the corresponding $P^+$ eigenvectors,
% \begin{subequations}
\begin{align}
\bm{v}_1^{-}=\begin{pmatrix}
0.843 \\ 
0.448 \\ 
0.298
\end{pmatrix}, 
 \; \; \; 
 \bm{v}_2^{-}=\begin{pmatrix}
0 \\ 
0.432 \\ 
-0.902
\end{pmatrix},
 \; \; \;  \bm{v}_3^{-}=\begin{pmatrix}
0 \\ 
0.928 \\ 
0.374
\end{pmatrix}.
\end{align}
%\end{subequations}
Since all three eigenvalues are positive
both fixed points $P^+$ and $P^-$ are tricritical.
The eigendirections are indicated by red arrows in 
Fig.~\ref{fig:RGflowd4}.
For completeness, we also show
in Fig.~\ref{fig:RGflow4} the projections of the RG flow in $d=4$
%where we have the Pomeranchuk fixed points,
onto the $\tilde{g}$ -$ \tilde{h}$ plane at $\tilde{r}_l=\tilde{r}_\ast$ and the $\tilde{r}$-$\tilde{h}$ plane at $\tilde{g}_l=\tilde{g}^+_\ast$ and $\tilde{g}_l=\tilde{g}^-_\ast$, respectively.
Recall that the  relevant coupling $\tilde{h}_l$ is related to the renormalized one-point vertex
of our effective phonon action and is related to the shift in the chemical potential
which is necessary to keep the density constant; see Eqs.~(\ref{eq:onepoint1}).
Note that $\tilde{h}_l$ is exactly  analogous to an external magnetic field in the
Ginzburg-Landau-Wilson Hamiltonian for the Ising model \cite{LeBellac91}.
The existence of a pair of nontrivial ultraviolet-stable fixed points
in our effective phonon theory for the Holstein model
defined by the Euclidean action $S_{\rm eff} [ X ]$
in Eq.~(\ref{eq:Seffphonon}) is closely related to a similar pair of fixed points
of scalar $\phi^3$-theory above six dimensions \cite{Polyakov70,Migdal71,Mack73,Fisher78}. 
The shift in the dimensionality is due to the fact that
the quantum action  $S_{\rm eff} [ X ]$ is characterized by
a dynamical exponent $z=3$ so that the critical  dimension above which these fixed points
 emerge is reduced from six to
$6 -z =3$, see Ref.~[\onlinecite{Hertz76}]. 
Note that in the context of $\phi^3$-theory the ultraviolet-stable 
nontrivial fixed points above six dimensions have recently received renewed
interest in high-energy physics \cite{Fei14,Rong20} because 
these fixed points offer a possibility 
to construct a well-defined continuum limit of perturbatively nonrenormalizable field theories 
(asymptotic safety \cite{Weinberg79,Reuter98}).

\begin{figure}
\begin{center}
  \centering
\vspace{7mm}
\includegraphics[width=0.3\textwidth]{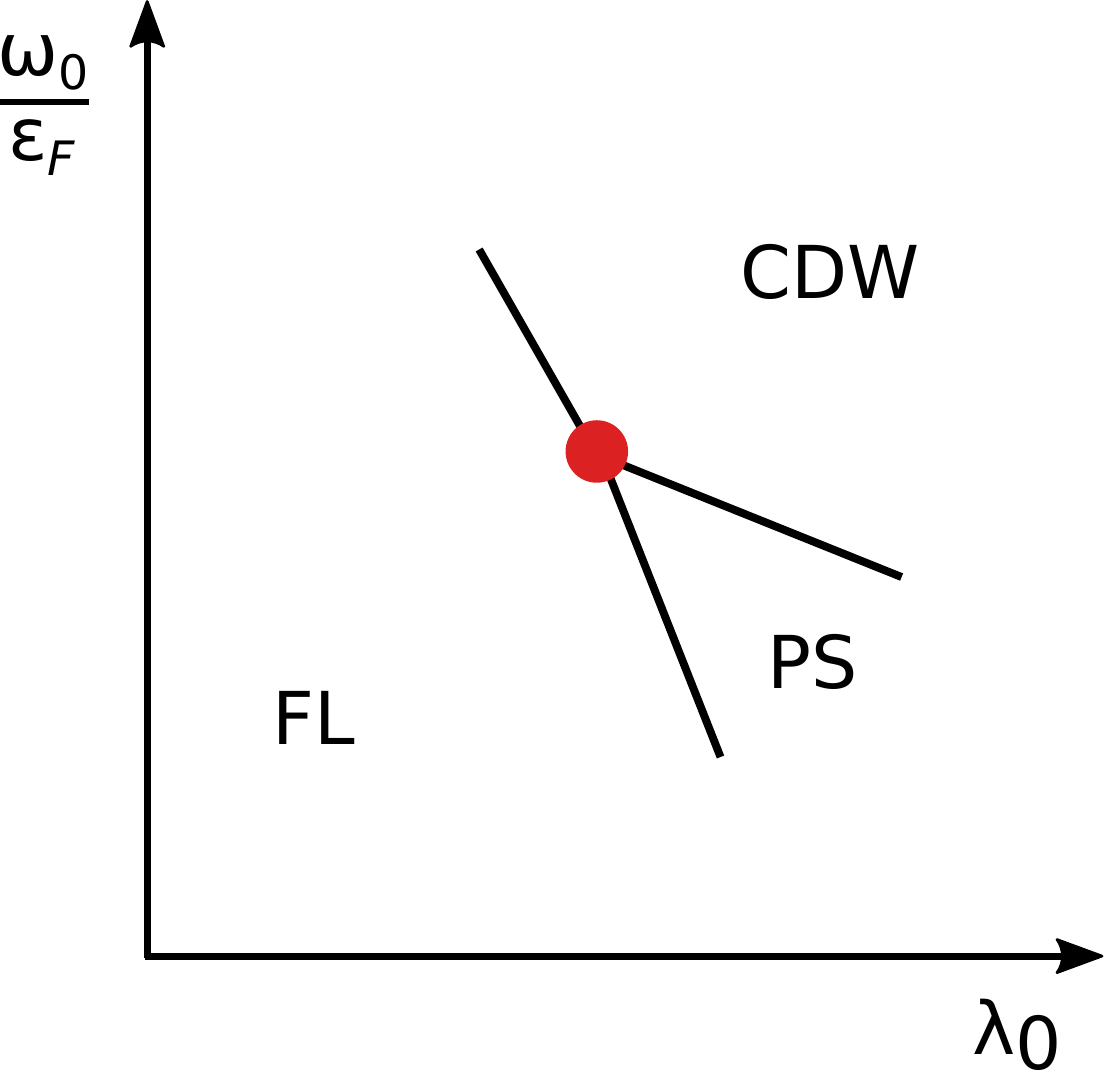}%\nspace
   \end{center}
%  \vspace{-4mm}
  \caption{%
%(Color online)
Schematic ground state phase diagram of the Holstein model in $d=3$  
in the plane spanned by the electron-phonon coupling $\lambda_0$ and the adiabatic 
ratio $\omega_0/\epsilon_F$ for fixed but not too small densities. 
We have ignored a possible superfluid phase, assuming that 
the weak-coupling phase is a normal Fermi liquid (FL).
In the  adiabatic regime $\omega_0 / \epsilon_F \ll 1$ we expect a first-order transition to a 
phase-separated (PS) inhomogeneous state when $\lambda_0$
exceeds a certain threshold of order unity. 
However, in the antiadiabatic regime $\omega_0 / \epsilon_F \gg 1$ there is
a first-order transition to a state with  charge-density-wave (CDW) order.
In $d \leq 3$ the intersection of the three phase boundaries (red dot) 
is a noncritical triple point.
For $d > 3$ the triple point transforms into a critical point 
while the phase boundary between the FL and the PS phases remains first order. 
}
\label{fig:Phasediagram}
\end{figure}

Next, let us discuss the implications of our RG analysis for the
phase diagram of the Holstein model.
From the flow diagrams in Figs.~\ref{fig:RGflow123}, \ref{fig:RGflowd4}, and \ref{fig:RGflow4}
it is clear that
in all dimensions
the RG flow of the Holstein model in the  $\tilde{g}$ - $\tilde{r}$-plane
is divided by a separatrix into two regimes.
In the first regime $\tilde{r}_l$ flows to positive values so that the 
renormalized phonon frequency is finite. 
Since we do not take into account possible superfluid states,
the ground state is then expected to be a normal Fermi liquid.
In the second regime
$\tilde{r}_l$ flows to negative values and eventually diverges at 
a finite scale $\Lambda_*$. The interpretation of this runaway flow is
somewhat ambiguous so that we need information from other methods to draw conclusions
for the phase diagram.
In two dimensions various numerical calculations \cite{Kumar08,Ohgoe17,Esterlis18,Esterlis19,Wang20} found that 
for sufficiently strong electron-phonon coupling
the Holstein model exhibits a first-order transition
from a  Fermi liquid (FL) phase to a charge-density-wave (CDW) phase.
In addition, some authors \cite{Ohgoe17} also found
that in a certain interval of densities and for not too large values of the adiabatic ratio $\omega_0 / \epsilon_F$
the system phase separates (PS) into regions with different densities. 
The transition between the FL and CDW as well as the transition
between the FL and the PS phase are first order.
The fact that in the antiadiabatic limit $ \omega_0 \gg \epsilon_F$ the
FL phase becomes unstable towards a CDW  can also understood by considering
the effective electronic action obtained by integrating over the phonons,
 \begin{equation}
 S_{\rm eff} [ \bar{c} ,c ] = - \int_K G_0^{-1} ( K ) \bar{c}_K c_K -
 \frac{ \gamma_0^2}{2} \int_Q D_0 ( Q ) \rho_{-Q} \rho_Q ,
 \label{eq:Seffc}
 \end{equation}
where $\rho_Q = \int_K \bar{c}_K c_{ K+Q }$ represents the electronic density.
In the antiadiabatic limit the free phonon propagator can be  approximated by $D_0 (Q ) \approx 1/ \omega_0^2$
so that the last term in Eq.~(\ref{eq:Seffc}) reduces to a local attractive interaction with strength
$- \gamma_0^2 / \omega_0^2 = - \lambda_0 / \nu$. For not too small densities
the dominant instability of this model is expected to be CDW \cite{Wang20,Grzybowski07}.
Assuming that the phase diagram of the Holstein model
in $d=3$ is not qualitatively different from the
phase diagram in $d=2$ (this assumption is supported by 
Fig.~\ref{fig:RGflow123} which
shows  qualitatively similar RG flows in $d=3$ and $d=2$),
we propose that in three dimensions the phase diagram of the Holstein model 
as a function of the electron-phonon coupling $\lambda_0$ and the adiabatic ratio
$\omega_0 / \epsilon_F$ has for fixed, but  not too small, densities
the form sketched in  Fig.~\ref{fig:Phasediagram}.
Here all phase boundaries are first order and the intersection of the
three phase boundaries is not a critical point but a triple point because
for $d \leq 3$ our RG analysis rules out a
critical point with a divergent
uniform compressibility.

However, for $d > 3$ we expect that the intersection 
of the three phase boundaries in Fig.~\ref{fig:Phasediagram} becomes a critical point.
The critical behavior close to this point is then controlled by one of the  Pomeranchuk
fixed points discussed above. The phase boundary between the
FL and the PS phase is still first order, because
we know that both the adiabatic ratio $\omega_0 / \epsilon_F$ and the electron-phonon coupling
$\lambda_0$ have to be fine-tuned to
obtain criticality.
The nature of the other phase boundaries remains to be investigated, but this is beyond the scope of this work.

\section{Summary and conclusions}

In this work we have presented evidence that 
in dimensions $d > 3$ the Holstein model exhibits a quantum critical point associated with
a Pomeranchuk instability. The underlying critical RG fixed point is ultraviolet stable and
can only be realized if both the dimensionless electron-phonon coupling
$\lambda_0$ and the adiabatic ratio $\omega_0 / \epsilon_F$ are fine-tuned.
This fixed point is closely related to the
well-known ultraviolet-stable fixed point of $\phi^3$-theory above six dimensions;
the dynamic exponent $z=3$ reduces the relevant critical dimension from six to three.
Note that the bare value of the relevant three-point vertex in our effective phonon action
is proportional to the third power 
$\lambda_0^3$ of dimensionless electron-phonon coupling so that second-order perturbation theory in $\lambda_0$ is not sufficient to detect the singularities 
associated with the dominant critical fluctuations.
Our result that for  $d \leq 3$ the RG flow of Holstein model does not have a 
nontrivial fixed point associated with a Pomeranchuk instability does not exclude the possibility that the phase diagram contains 
regimes where the state of the system exhibits phase separation.
However, the transitions to this state must be first order.
By combining our RG analysis with numerical results for the phase diagram of the two-dimensional Holstein model
by other authors \cite{Kumar08,Ohgoe17,Esterlis18,Esterlis19,Wang20} we propose the schematic
phase diagram of the Holstein model shown in Fig.~\ref{fig:Phasediagram}.

For our calculation of the compressibility we have used the exact 
Ward identity (\ref{eq:wiphon}) to express the compressibility in terms of the renormalized phonon frequency of the Holstein model.
For our purpose, it is therefore sufficient to analyze the effective phonon action $S_{\rm eff} [ X ]$
defined in Eq.~(\ref{eq:Seffphonon}). Unfortunately, the renormalized 
electron-phonon vertex cannot be calculated within this approach so that
we cannot make statements 
about the validity of Migdal's theorem close the Pomeranchuk instability.
In principle our FRG approach can be generalized to obtain also flow equations 
for the electron-phonon vertices, but this is beyond the scope of this work.

Finally, let us comment on the significance of our finding that the Pomeranchuk fixed points
of the Holstein model in $d > 3$ are ultraviolet stable.
RG fixed points with this property are crucial to define a well-defined continuum limit
in perturbatively nonrenormalizable field theories. 
If these fixed points are non-Gaussian, the theory is called asymptotically safe, a prominent candidate  being quantum gravity \cite{Weinberg79}, where  evidence for 
the existence of a nontrivial 
ultraviolet-stable fixed point has been obtained by means of FRG methods \cite{Reuter98,Dupuis21}.
In this sense, the effective phonon action of the Holstein model in $d > 3$ defines an asymptotically save field
theory.
However, for $d  \leq 3$ the ultraviolet-stable fixed point of the Holstein model is Gaussian
so that the
interaction vanishes at the fixed point (asymptotic freedom).
Of course, the effective phonon action of the Holstein model 
has an intrinsic ultraviolet cutoff given by $q_0$ defined in Eq.~(\ref{eq:q0def}), 
but for all RG trajectories  which emanate from the UV-stable fixed points
the UV cutoff can be removed without affecting any physical observables.

\section*{ACKNOWLEDGMENTS}

Most of this work was completed during a sabbatical stay at the
Department of Physics and Astronomy at the University of California, Irvine.
% during which N.C. and M.H. had the opportunity to visit. 
The authors thank Sasha Chernyshev for his hospitality. We are also grateful to the Deutsche Forschungsgemeinschaft (DFG, German Research Foundation) for financial support via  TRR 288 - 422213477 (Project A07).

\begin{appendix}

\section*{APPENDIX A:  
Dyson-Schwinger equations and Ward identities for the Holstein model}
%including next-nearest-neighbor exchange}

\setcounter{equation}{0}
\renewcommand{\theequation}{A\arabic{equation}}

The renormalized phonon frequency in the Holstein model
is related to the compressibility via the Ward identity  (\ref{eq:wiphon})
which is
closely related to the 
well-known compressibility sum 
rule \cite{Pines89,Kubo04,Guo13}.
The specific  form of this Ward identity
for the Holstein model apparently cannot be found in the literature.
In this Appendix we therefore give a self-contained derivation of Eq.~(\ref{eq:wiphon}) and
related Ward identities using functional methods.

\subsection{Dyson-Schwinger equations}

To begin with, we derive
Dyson-Schwinger equations (also called skeleton equations) for the Holstein model,
relating correlation functions of different order. 
Therefore
we follow the method outlined
in Refs.~[\onlinecite{Kopietz10,Schuetz05}]. Consider the generating functional of the
Euclidean correlation functions of the Holstein model,
 \begin{equation}
 {\cal{G}} [ \bar{\eta} , \eta , J ] = \int
 {\cal{D}} [ \bar{c} , c , X ] e^{ - S + ( \bar{\eta} , c ) + ( \bar{c} , \eta ) + ( J , X ) },
 \label{eq:funcint}
 \end{equation}
where $\bar{\eta}$ and $\eta$ are Grassmann sources, $J$ is a bosonic source field,
and we have used the abbreviations
 \begin{eqnarray}
  ( \bar{\eta} , c ) + ( \bar{c} , \eta ) & = & \int_K ( \bar{\eta}_K c_K + \bar{c}_K \eta_K ),
 \\
 ( J , X ) & = & \int_Q J_{-Q} X_Q.
 \end{eqnarray}
The Euclidean bare action $S [ \bar{c} , c , X ]$ of the Holstein model
is given in Eq.~(\ref{eq:ScX}).
Using the invariance of the functional integral in Eq.~(\ref{eq:funcint}) with respect to 
infinitesimal shifts in the integration variables
we obtain the functional equations
 \begin{subequations}
 \label{eq:DSv1}
 \begin{eqnarray}
 \left( J_{-Q} - D_0^{-1} (Q) \frac{\delta}{\delta J_Q } \right) {\cal{G}} - \zeta \gamma_0 
\int_K \frac{ \delta^2 {\cal{G}}}{ \delta \eta_{ K+Q} \delta \bar{\eta}_K } & = & 0,
 \nonumber
 \\
 & &
 \\
\left( \zeta \bar{\eta}_{K} - \bar{G}_0^{-1} (K) \frac{\delta}{\delta \eta_K } \right) {\cal{G}} -  \gamma_0 
\int_Q \frac{ \delta^2 {\cal{G}}}{ \delta \eta_{ K+Q} \delta J_{-Q} } & = & 0,
 \nonumber
 \\
 & &
 \\
\left( {\eta}_{K} - \bar{G}_0^{-1} (K) \frac{\delta}{\delta \bar{\eta}_K } \right) {\cal{G}} -  \gamma_0 
\int_Q \frac{ \delta^2 {\cal{G}}}{ \delta \bar{\eta}_{ K-Q} \delta J_{-Q} } & = & 0,
 \nonumber
 \\
 & &
 \end{eqnarray} 
\end{subequations}
where $\bar{G}_0 ( K ) = - G_0 ( K )$ and 
we have introduced the fermionic statistics factor $\zeta =-1$. Next, we express the above Dyson Schwinger equations in terms of
 the generating functional of connected correlation functions
 \begin{equation}
  {\cal{G}}_c [ \bar{\eta} , \eta , J ] = \ln    {\cal{G}} [ \bar{\eta} , \eta , J ] ,
 \end{equation}
and its subtracted Legendre transform
  \begin{eqnarray}
& &  \Gamma [ \bar{\psi} , \psi , \phi ]  =   ( \bar{\eta} , \psi ) + ( \bar{\psi} , \eta ) + ( J , \phi ) - {\cal{G}}_c [ \bar{\eta} , \eta , J ] 
 \nonumber
 \\
 & & - \int_K \bar{G}_0^{-1} ( K ) \bar{\psi}_K \psi_K -
 \frac{1}{2} \int_Q D_0^{-1} ( Q)  \phi_{ - Q } \phi_Q,
 \end{eqnarray}
where on the right-hand side the sources should be expressed in terms of the field expectation values by inverting the relations
 \begin{subequations} 
 \begin{eqnarray}
 \frac{ \delta {\cal{G}}_c }{ \delta \bar{\eta}_K } & = & 
 \psi_K = \langle c_K \rangle,
 \\
 \frac{ \delta {\cal{G}}_c }{ \delta {\eta}_K } & = & \zeta 
 \bar{\psi}_K = \zeta \langle \bar{c}_K \rangle,
 \\
  \frac{ \delta {\cal{G}}_c }{ \delta J_{-Q} } & = & 
 \phi_{Q} = \langle X_{Q} \rangle.
 \end{eqnarray}
 \end{subequations}
Note that by construction
  \begin{subequations} 
 \begin{eqnarray}
 \frac{ \delta \Gamma }{ \delta \bar{\psi}_K }   + \bar{G}_0^{-1} ( K ) \psi_K    & = & 
 \eta_K,
 \\
 \frac{ \delta \Gamma }{ \delta {\psi}_K } +  \zeta \bar{G}_0^{-1} ( K ) \bar{\psi}_K   & = & \zeta 
 \bar{\eta}_K ,
 \\
  \frac{ \delta \Gamma }{ \delta \phi_{-Q} }  + D_0^{-1} ( Q )  \phi_{Q}   & = & 
 J_{Q}.
 \label{eq:Gammaphi}
 \end{eqnarray}
 \end{subequations}
The Dyson-Schwinger equations (\ref{eq:DSv1}) then reduce to
 \begin{subequations}
 \label{eq:DysonSchwinger}
 \begin{eqnarray}
 \frac{ \delta \Gamma}{ \delta \phi_Q} -
 \gamma_0 \int_K \left(
 \bar{\psi}_{ K+Q} \psi_K + \frac{ \delta^2 {\cal{G}}_c }{\delta \bar{\eta}_K
 \delta \eta_{ K + Q}}\right) & = & 0,
 \label{eq:DS1}
 \hspace{7mm}
 \\
  \frac{ \delta \Gamma}{ \delta \psi_K } -
 \gamma_0 \int_Q \left(
 \zeta \bar{\psi}_{ K+Q} \phi_Q + \frac{ \delta^2 {\cal{G}}_c }{\delta 
 {\eta}_{K+Q}
 \delta J_{ - Q}}\right) & = & 0,
 \hspace{10mm}
 \\
  \frac{ \delta \Gamma}{ \delta \bar{\psi}_K } -
 \gamma_0 \int_Q \left(
 {\psi}_{ K-Q} \phi_Q + \frac{ \delta^2 {\cal{G}}_c }{\delta 
 \bar{\eta}_{K-Q}
 \delta J_{ - Q}}\right) & = & 0.
 \label{eq:DS3}
 \end{eqnarray}
 \end{subequations}
By taking successive derivatives of Eqs.~(\ref{eq:DysonSchwinger})
with respect to the field expectation values and then setting the sources equal to zero 
we obtain an infinite set of  Dyson-Schwinger equations for the irreducible vertices.

First of all, let us consider Eq.~(\ref{eq:DS1}) for vanishing sources,
taking into account that for finite density
the expectation value $\phi_Q$ of the phonon field
has a finite limit $\phi_Q^0$.
Here the superscript means that
$\phi^0_Q = \langle X_Q \rangle$ is calculated for vanishing sources $\eta = \bar{\eta} = J =0$. Using Eq.~(\ref{eq:Gammaphi}) we obtain
 \begin{eqnarray}
 D_0^{-1} ( - Q ) \phi^0_{-Q} & = &  - \gamma_0 \int_K 
 \left. \frac{ \delta^2 {\cal{G}}_c }{\delta \bar{\eta}_K
 \delta \eta_{ K + Q}}\right|_{ \eta = \bar{\eta} = J =0}
 \nonumber
 \\
 & = & - \gamma_0 \delta ( Q) \int_K G ( K ).
 \label{eq:DSphi0a}
 \end{eqnarray}
The integral $\int_K G ( K ) = \rho$ can be identified with the  
exact electronic density of the system
so that we can write
 \begin{equation}
 \phi^0_Q = \delta ( Q) {\phi^0}, \; \; \; \; \; \; \phi^0  = - \frac{\gamma_0}{\omega_0^2}  \rho.
 \label{eq:DSphi0}
 \end{equation}
We conclude that for any finite electronic density the phonon displacement
field of the Holstein model has a finite expectation value.

\begin{figure}[tb]
 \begin{center}
  \centering
\includegraphics[width=0.47\textwidth]{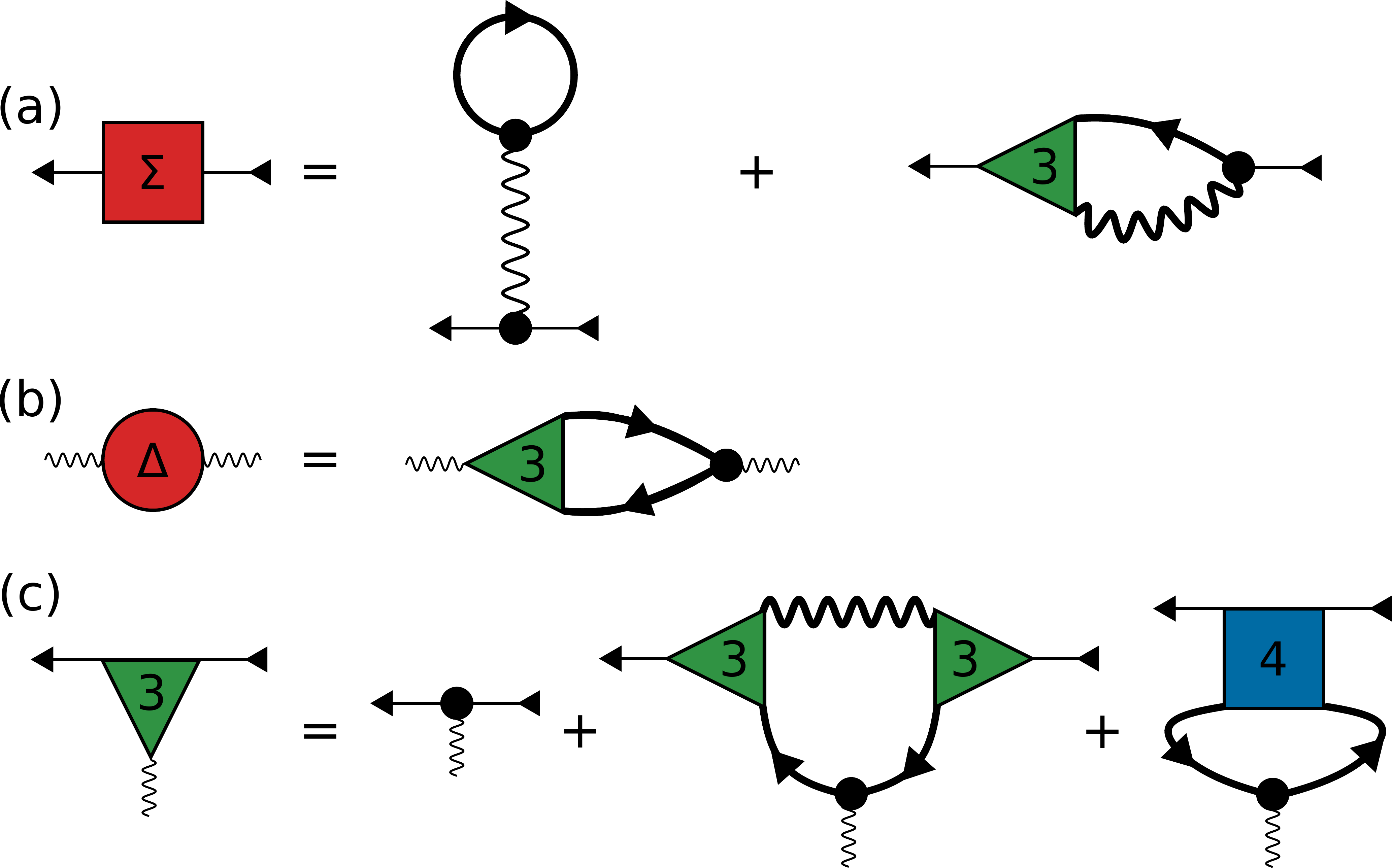}%\nspace
   \end{center}
%  \vspace{-4mm}
  \caption{%
%(Color online)
Diagrammatic representation 
of Dyson-Schwinger equations for the Holstein model: (a)  
electronic self-energy, (b) phonon self-energy,  and (c) electron-phonon vertex 
(represented by a green triangle).
The solid arrows represent the exact electron propagator $G ( K )$ and the thick
wavy lines represent the exact phonon propagator $D ( Q )$. The thin wavy line in the tadpole contribution to the
electronic self-energy represents the bare static 
phonon propagator $D_0 (0) = 1/ \omega_0^2$.
The bare phonon vertex $\gamma_0$ is represented by a black dot and  
the irreducible fermionic 
two-body interaction  $\Gamma^{ \bar{c} \bar{c} cc} ( K + Q , K^{\prime}  , K^{\prime} + Q  , K )$ is represented by a blue square.
}
\label{fig:DS}
\end{figure}

Next, we derive the Dyson-Schwinger equation for the electronic self-energy.
Applying $\frac{\delta}{\delta \psi_K^{\prime} }$ to both sides of
Eq.~(\ref{eq:DS3}) and then setting 
all sources equal to zero gives
 \begin{eqnarray}
 & &  \delta ( K - K^{\prime} ) \Sigma ( K )  =   \gamma_0 \phi^0_{K - K^{\prime} }
 \nonumber
 \\
 &  & +  \gamma_0 \left. \int_Q \frac{ \delta^3 {\cal{G}}_c}{ \delta \psi_{ K^{\prime}}
 \delta \bar{\eta}_{ K-Q} \delta J_{-Q}} \right|_{ \eta = \bar{\eta} = J =0}.
 \end{eqnarray}
The last term can be expressed in terms of irreducible vertices as follows \cite{Kopietz10,Schuetz05},
 \begin{eqnarray}
 & &  \left.  \frac{ \delta^3 {\cal{G}}_c}{ \delta \psi_{ K^{\prime}}
 \delta \bar{\eta}_{ K-Q} \delta J_{-Q}} \right|_{ \eta = \bar{\eta} = J =0} =
 \delta ( K-K^{\prime} ) G ( K -  Q )
 \nonumber
 \\
 &  &   \hspace{20mm} \times \Gamma^{\bar{c} c \varphi}
 ( K -  Q, K , - Q )  D ( - Q )  , \hspace{7mm}
 \end{eqnarray} 
where the three-legged vertex $\Gamma^{\bar{c} c \varphi}
 ( K + Q, K , Q )$ is defined by expanding the functional
$\Gamma [ \bar{\psi} , \psi , \phi ]$ around $ \phi = \phi^0$, i.e.,
 \begin{eqnarray}
  & & \Gamma [ \bar{\psi} , \psi , \phi^0 + \varphi ]  =  \Gamma [ 0,0 , \phi^0 ] 
 - \gamma_0 \rho \varphi_{ Q=0}
 \nonumber
 \\
 &  & +  \int_K \Sigma ( K ) \bar{\psi}_K \psi_K +   \frac{1}{2} \int_Q \Delta ( Q ) \varphi_{-Q} \varphi_Q 
 \nonumber
 \\
 &  & + \int_K \int_Q \Gamma^{ \bar{c} c \varphi} ( K + Q , K , Q )
 \bar{\psi}_{ K+Q} \psi_K \varphi_Q + \ldots \; .
 \hspace{7mm}
 \end{eqnarray}
Note that the vertex expansion of the Legendre transform
  \begin{eqnarray} 
& &  {\cal{L}} [ \bar{\psi} , \psi , \phi ]  =    ( \bar{\eta} , \psi ) + ( \bar{\psi} , \eta ) + ( J , \phi ) - {\cal{G}}_c [ \bar{\eta} , \eta , J ] 
 \nonumber
 \\
 &  & = {\Gamma} [ \bar{\psi} , \psi , \phi ]     + \int_K \bar{G}_0^{-1} ( K ) \bar{\psi}_K \psi_K +
 \frac{1}{2} \int_Q D_0^{-1} ( Q)  \phi_{ - Q } \phi_Q
 \nonumber
 \\
 & &
 \end{eqnarray}
does not have a linear term,
\begin{eqnarray}
  & & {\cal{L}} [ \bar{\psi} , \psi , \phi^0 + \varphi ]  =  {\cal{L}} [ 0,0 , \phi^0 ] 
 \nonumber
 \\
 &  & -   \int_K G^{-1} ( K ) \bar{\psi}_K \psi_K +   \frac{1}{2} \int_Q D^{-1} ( Q ) \varphi_{-Q} \varphi_Q 
 \nonumber
 \\
 &  & + \int_K \int_Q \Gamma^{ \bar{c} c \varphi} ( K + Q , K , Q )
 \bar{\psi}_{ K+Q} \psi_K \varphi_Q + \ldots \; .
 \hspace{7mm}
 \end{eqnarray}
Here the exact electron and phonon propagators can be expressed via the corresponding self-energies $\Sigma ( K )$ and $ \Delta ( Q )$ via the Dyson equations
 \begin{eqnarray}
 G ( K ) & = & \frac{1}{ i \omega - \epsilon_{\bd{k}} + \mu - \Sigma ( K ) },
 \\
 D ( Q ) & = & \frac{1}{ \bar{\omega}^2 + \omega_0^2 + \Delta ( Q ) }.
 \end{eqnarray}
The Dyson-Schwinger equation for  the electronic self-energy can then be written as
 \begin{eqnarray}
 & &  \Sigma ( K )  =  \gamma_0 \phi^0 
 \nonumber
 \\
 &  &  + \gamma_0 \int_Q   D ( Q ) G ( K + Q )  \Gamma^{\bar{c} c \varphi}
 ( K + Q, K , Q )   ,
 \label{eq:DSSigma}
 \end{eqnarray}
which is shown diagrammatically in Fig.~\ref{fig:DS}  (a).
The first term  
 \begin{equation}
 \Sigma_\rho = \gamma_0 \phi^0  = - \gamma_0^2 D_0 (0) \rho = 
 - \frac{ \gamma_0^2 }{\omega_0^2} \rho
 \end{equation}
can be identified with
the sum of all tadpole contributions to the irreducible electronic self-energy.
To leading order in the interaction, this is the Hartree correction.
The negative sign of the Hartree self-energy can be simply understood from
the fact the effective interaction
between the electrons mediated by the phonons is attractive for small momentum and energy transfers. To see this, we integrate the exponentiated Euclidean action $e^{- S}$ over the phonon field to obtain that the effective electronic action of the Holstein model, given in Eq.~(\ref{eq:Seffc}),
% \begin{equation}
% S_{\rm eff} = - \int_K G_0^{-1} ( K ) \bar{c}_K c_K - \frac{1}{2} \int_Q \gamma_0^2 D_0 (Q) \rho_{-Q} \rho_Q,
% \label{eq:Seff}
% \end{equation}
has an attractive two-body interaction 
 $ - \gamma_0^2 D_0 (0) = 
- \gamma_0^2/ \omega_0^2$ in the limit of vanishing energy-momentum transfer $Q=0$.

Next, by taking the derivative $\frac{\delta}{\delta \phi_{-Q} }$
of Eq.~(\ref{eq:DS1}) and then taking the limit of vanishing sources
we obtain the Dyson-Schwinger equation for the phonon self-energy,
 \begin{equation}
 \Delta (Q) = \gamma_0 \int_K  G ( K ) G ( K+ Q ) \Gamma^{ \bar{c} c \varphi } ( K + Q , K , Q )  ,
 \label{eq:DSW}
 \end{equation}
which is shown diagrammatically in Fig.~\ref{fig:DS} (b).
Finally, applying $\frac{ \delta^2}{ \delta \bar{\psi}_{ K+ Q} \delta \psi_K}$ to both sides of  Eq.~(\ref{eq:DS1}) and then setting the sources equal to zero we obtain the Dyson-Schwinger equation for the electron-phonon vertex
 \begin{widetext}
 \begin{eqnarray}
 \Gamma^{ \bar{c} c \varphi} ( K + Q , K , Q )  
%& = &   \gamma_0 
%  + \gamma_0
% \int_{K^{\prime}} G ( K^{\prime} ) G ( K^{\prime} + Q )
% \Gamma^{ \bar{c} \bar{c} cc} ( K + Q , K^{\prime}  , K^{\prime} + Q  , K )
% \nonumber
% \\
% & + &  \gamma_0 \int_{ Q^{\prime}} \Gamma^{ \bar{c} c \varphi }
% ( K + Q^{\prime} , K , Q^{\prime} ) G ( K + Q^{\prime} ) G ( K + Q +  Q^{\prime}  ) 
%  \Gamma^{ \bar{c} c \varphi } ( K + Q , K + Q + Q^{\prime}, - Q^{\prime} ) D ( Q^{\prime} )
% \nonumber
% \\
 & = & \gamma_0 
  + \gamma_0
 \int_{K^{\prime}} G ( K^{\prime} ) G ( K^{\prime} + Q )
 \Bigl\{
 \Gamma^{ \bar{c} \bar{c} cc} ( K + Q , K^{\prime}  , K^{\prime} + Q  , K )
 \nonumber
 \\
 &  & \hspace{16mm} +  
 \Gamma^{ \bar{c} c \varphi } ( K + Q , K^{\prime} + Q , K - K^{\prime} ) 
 D ( K^{\prime} - K )  
 \Gamma^{ \bar{c} c \varphi } ( K^{\prime} , K , K^{\prime} - K  ) 
 \Bigr\},
 \label{eq:DSGamma}
\end{eqnarray}
 \end{widetext}
which is shown diagrammatically in Fig.~\ref{fig:DS} (c).
Here  $\Gamma^{ \bar{c} \bar{c} cc} ( K + Q , K^{\prime}  , K^{\prime} + Q  , K )$
is the effective two-body interaction between the fermions which is irreducible with respect to cutting a single electron line or a single phonon line.

\subsection{Ward identities}

We now show that for $Q =0$ the phonon self-energy 
can be expressed in terms of the compressibility
$\partial \rho / \partial \mu$ via the Ward identity
 \begin{equation}
 \Delta (0) = - \frac{ \gamma_0^2 \frac{\partial \rho}{\partial \mu } }{
 1 + \frac{ \gamma_0^2}{\omega_0^2} \frac{\partial \rho}{\partial \mu} }.
 \label{eq:Warddelta}
 \end{equation}
For $Q=0$ this implies that the inverse phonon propagator is given by
 \begin{equation}
 D^{-1} (0) \equiv \tilde{\omega}_0^2 =  \omega_0^2 + \Delta ( 0 ) = \frac{\omega_0^2}{
 1 + \frac{ \gamma_0^2}{\omega_0^2} \frac{\partial \rho}{\partial \mu} },
 \label{eq:phonren}
 \end{equation}
which is the Ward identity (\ref{eq:wiphon}).
Note that  thermodynamic stability implies that
the compressibility is nonnegative so that 
$D^{-1} (0 ) \geq 0$.
The identities (\ref{eq:Warddelta}) and (\ref{eq:phonren}) 
imply that the phonon self-energy $\Delta (0)$ of the Holstein model is 
strictly negative as long as the normal state is thermodynamically stable. 
At the same time the renormalized phonon energy
is strictly positive and vanishes only when the compressibility diverges.

The identity (\ref{eq:Warddelta}) is closely related to the fact that
for $Q=0$ (more precisely: in the so-called $\bd{q}$-limit
where we first set $\bar{\omega} =0$ and then take the limit
$\bd{q} \rightarrow 0$) 
the exact electron-phonon vertex
 $\Gamma^{ \bar{c} c \varphi} ( K  , K , 0 )$ satisfies the following Ward identity,
 \begin{equation}
 \frac{\Gamma^{\bar{c} c \varphi} ( K , K , 0 ) }{ \gamma_0} =
 \frac{ 1 - \frac{\partial \Sigma ( K ) }{ \partial \mu}}{ 1 - \frac{ \partial \Sigma_\rho}{\partial \mu } }
 = \frac{ 1 - \frac{\partial \Sigma ( K ) }{ \partial \mu}}{ 1 + \frac{ \gamma_0^2}{\omega_0^2}
 \frac{ \partial \rho}{\partial \mu } }.
% =  1 - \frac{ \frac{ \partial [ \Sigma ( K ) - \Sigma_{\rho} ]}{\partial \mu }}{ 1 - \frac{\partial \Sigma_{\rho}}{\partial \mu } } .
 \label{eq:WIcomp}
 \end{equation}
Note that if we approximate the 
self-energy $\Sigma ( K )$ by its tadpole contribution $\Sigma_{\rho}$ the vertex $\Gamma^{\bar{c} c \varphi} ( K , K , 0 )$ 
is not renormalized.

To prove the Ward identity (\ref{eq:WIcomp}), we use the
exact FRG flow equation for the fermionic self-energy  in the 
chemical potential cutoff scheme \cite{Sauli06}, which 
for the Holstein model is
given by
 \begin{equation}
 \frac{ \partial \Sigma ( K )}{\partial \mu }  =  
 \Gamma^{\bar{c} c \varphi} ( K, K , 0 )
 \frac{\partial \phi^0}{\partial \mu }  - I ( K ),
 \label{eq:Sigmamuflow}
 \end{equation}
with
 \begin{eqnarray}
 & & I ( K )   =   \int_{K^{\prime}} G^2 ( K^{\prime} )
 \Bigl\{
 \Gamma^{ \bar{c} \bar{c} cc} ( K , K^{\prime}  , K^{\prime}   , K )
 \nonumber
 \\
 &  &  +       \Gamma^{ \bar{c} c \varphi }
 ( K  , K^{\prime}, K - K^{\prime} )  D ( K^{\prime} - K  )
% \nonumber
% \\
% & & \hspace{4mm} \times 
\Gamma^{ \bar{c} c \varphi }
 ( K^{\prime} , K , K^{\prime} - K ) \Bigr\} .
 \nonumber
 \\
 & &
 \label{eq:Iflowdef}
\end{eqnarray}
However, for $Q=0$ the Dyson-Schwinger equation
(\ref{eq:DSGamma}) for the electron-phonon vertex can be written as
\begin{equation}
 \Gamma^{ \bar{c} c \varphi} ( K , K , 0 ) = \gamma_0 + \gamma_0 I ( K ).
 \end{equation}
Using this to eliminate  $I ( K )$ in the flow equation (\ref{eq:Sigmamuflow})
we obtain
 \begin{equation}
 \Gamma^{ \bar{c} c \varphi} ( K , K , 0 ) = \gamma_0 + \gamma_0 \left[ \Gamma^{ \bar{c} c \varphi} ( K , K , 0 )
 \frac{\partial \phi^0}{\partial \mu} - \frac{ \partial \Sigma ( K ) }{\partial \mu} \right].
 \end{equation}
Solving for $\Gamma^{ \bar{c} c \varphi} ( K , K , 0 )$ and noting that
$\gamma_0 \phi^0 = \Sigma_\rho $ we obtain the Ward identity~(\ref{eq:WIcomp}).
This identity implies the so-called compressibility sum rule \cite{Pines89},
which for the Holstein  model has the form
 \begin{equation}
 \frac{\partial \rho}{\partial \mu}  \equiv \frac{\partial}{\partial \mu} \int_K G ( K ) =
 \frac{ \Pi ( 0 )}{ 1 -  \gamma_0^2 D_0 (0) \Pi ( 0 ) },
 \label{eq:compsum}
 \end{equation}
where $\Pi (0) = \lim_{\bd{q}  \rightarrow 0} \Pi ( \bd{q} , i \omega =0 )$ is the so-called $\bd{q}$-limit
of the irreducible polarization $\Pi ( Q )$. The latter satisfies the
Dyson-Schwinger equation
 \begin{equation}
 \Pi ( Q) = - \gamma_0^{-1} \int_K  G ( K) G ( K + Q )  \Gamma^{ \bar{c} c \varphi} ( K+Q, K , Q )
 \label{eq:PiDS}
 \end{equation}
and is related to the phonon self-energy via
 \begin{equation}
 \label{eq:DeltaPiconnection}
 \Delta ( Q) = - \gamma_0^2 \Pi ( Q ).
 \end{equation}
Setting $Q=0$ in the Dyson-Schwinger equation (\ref{eq:PiDS}) and substituting the result into
Eq.~(\ref{eq:compsum}) we then obtain
 \begin{eqnarray}
 \frac{\partial \rho}{\partial \mu} & = & - \int_K \left( 1 - \frac{\partial \Sigma ( K ) }{\partial \mu} \right) G^2 (K) 
 \nonumber
 \\
 & = & -
  \left( 1 - \frac{ \partial \Sigma_\rho }{\partial \mu }  \right) \frac{1}{\gamma_0}
 \int_K \Gamma^{ \bar{c} c \varphi} ( K, K , 0 )  G^2 ( K ).
 \hspace{7mm}
 \label{eq:WIconsistent}
 \end{eqnarray}
The Ward identity (\ref{eq:WIcomp}) guarantees 
that this relation is indeed satisfied.
By inverting the chain of identities leading from Eq.~(\ref{eq:compsum}) to Eq.~(\ref{eq:WIconsistent}), we conclude that our Ward identity (\ref{eq:WIcomp}) is equivalent to the
compressibility sum rule~(\ref{eq:compsum}).

Finally, to proof the Ward identity (\ref{eq:Warddelta}), we 
multiply
 both sides of Eq.~(\ref{eq:compsum}) by $\gamma_0^2 D_0 (0)$ and use the fact that
 $
 \Sigma_\rho = - \gamma_0^2 D_0 (0) \rho
 $
is the tadpole contribution to the electronic self-energy. We then obtain
 \begin{equation}
 - \gamma_0^2 D_0 (0) \Pi (0) = \frac{ \frac{\partial \Sigma_\rho }{\partial \mu}}{
 1 -  \frac{\partial \Sigma_\rho }{\partial \mu} },
 \end{equation}
which can also be written as
 \begin{equation}
1 - \gamma_0^2 D_0 ( 0 ) \Pi (0) = \frac{1}{ 1 - \frac{ \partial \Sigma_\rho }{\partial \mu } },
 \label{eq:DPi}
 \end{equation}
and implies
 \begin{equation}
 \Delta ( 0 ) = \omega_0^2  \frac{ \frac{\partial \Sigma_\rho }{\partial \mu}}{
 1 -  \frac{\partial \Sigma_\rho }{\partial \mu} },
 \end{equation}
which is equivalent with Eq.~(\ref{eq:Warddelta}). Note that with the help of Eq.~(\ref{eq:DPi}) 
the Ward identity (\ref{eq:WIcomp}) for the electron-phonon vertex 
can alternatively  be written as
 \begin{eqnarray}
 \frac{\Gamma^{\bar{c} c \varphi} ( K , K , 0 ) }{ \gamma_0} & = &  \left[ 
 1 - \frac{\gamma_0^2}{\omega_0^2} \Pi (0) \right] \left[
  1 - \frac{\partial \Sigma ( K ) }{ \partial \mu} \right]
 \nonumber
 \\
 & = &  \frac{ \tilde{\omega}_0^2}{\omega_0^2}
 \left[
  1 - \frac{\partial \Sigma ( K ) }{ \partial \mu} \right].
 \label{eq:WIcomp2}
 \end{eqnarray}
This identity expresses the electron-phonon vertex at vanishing phonon momentum and energy 
in terms of the square of renormalized phonon frequency
$\tilde{\omega}_0^2 = \omega_0^2 + \Delta (0)$ and the derivative of the electronic self-energy
with respect to the chemical potential.

\section*{APPENDIX B:  
Symmetrized closed fermion loops}
%Leading interaction correction to the polarization}
%including next-nearest-neighbor exchange}
%
\setcounter{equation}{0}
\renewcommand{\theequation}{B\arabic{equation}}

The vertices in the effective phonon action $S_{\rm eff} [ X ]$ defined
in Eq.~(\ref{eq:SeffX}) can be expressed in terms of the symmetrized closed fermion loops $L_S^{(n)} ( Q_1 , \ldots , Q_n )$ as given by Eq.~(\ref{eq:gammalsym}).
The symmetrized closed fermion $n$-loop is defined by
 \begin{equation}
 L_S^{(n)} ( Q_1 , \ldots , Q_n ) = \frac{1}{n!}
 \sum_{ P ( 1, \ldots , n )} 
 L^{(n)} ( Q_{ P(1)}, \ldots , Q_{P (n)} ),
 \end{equation}
where the sum is over the $n!$ permutations of the labels and
$L^{(n)} ( Q_1 , \ldots , Q_n )$ is the corresponding nonsymmetrized loop. To define the latter, we  introduce shifted labels $\bar{Q}_j = \sum_{i=1}^{j-1} Q_i$, i.e.,
 \begin{eqnarray}
 \bar{Q}_1 & = & 0,
 \nonumber
 \\
 \bar{Q}_2 & = & Q_1,
 \nonumber
 \\
 \bar{Q}_3 & = & Q_1 + Q_2,
 \\
  &\vdots  &
 \nonumber
 \\
 \bar{Q}_n & = & Q_1 + \ldots + Q_{ n-1}.
 \nonumber
 \end{eqnarray}
The nonsymmetrized loop can then be written as
 \begin{equation}
 L^{(n)} ( Q_1 , \ldots , Q_n ) = \bar{L}^{(n)} ( \bar{Q}_1 , 
 \ldots , \bar{Q}_n ),
 \end{equation}
with
\begin{eqnarray}
& & \bar{L}^{(n)} ( \bar{Q}_1 , \ldots , \bar{Q}_n )  =    \int_K \prod_{ i=1}^n G_0 ( K - \bar{Q}_i )
 \nonumber
 \\
 &  & = \int_{\bd{k}}
 T \sum_{\omega} G_0 ( K - \bar{Q}_1 ) \cdots G_0 ( K - \bar{Q}_n ).
 \hspace{7mm}
 \end{eqnarray}
Here $\int_K = \int_{\bd{k}} T \sum_{\omega} $ and
$\int_{\bd{k}} = \int \frac{ d^d k}{(2 \pi )^d}$  denotes 
the $d$-dimensional momentum integration.
If we set all external momenta equal to zero, then we obtain \cite{Hertz74}
 \begin{equation}
 \bar{L}^{(n)} ( 0 , \ldots , 0 ) = \int_K [ G_0 ( K )  ]^n = 
\frac{1}{(n-1)!} \frac{ \partial^{n-1} 
 \rho_0 ( \mu  ) }{ \partial \mu^{n-1} },
 \end{equation}
where 
 \begin{equation}
 \rho_0 ( \mu ) = \int_K G_0 ( K ) = \int \frac{ d^d k}{ (2 \pi )^d}
 \frac{1}{ e^{ \beta ( \epsilon_{\bd{k}} - \mu ) } +1 }
 \end{equation}
is the density of noninteracting electrons as a function of the
chemical potential.
In particular, at zero temperature
 \begin{eqnarray}
 \bar{L}^{(2)} (0,0 ) & = & -\nu ( \mu ),
 \\
  \bar{L}^{(3)} (0,0 ,0) & = & \frac{1}{2} \frac{ \partial 
 \nu ( \mu )}{\partial \mu },\\
 & \vdots \nonumber\\
   \bar{L}^{(n)} (0,\ldots ,0) & = & \frac{(-1)^{n-1}}{(n-1)!} \frac{ \partial^{n-2}
    \nu ( \mu )}{\partial \mu^{n-2} },
 \end{eqnarray}
where
 \begin{equation}
 \nu ( \mu ) = \int \frac{ d^d k}{ ( 2 \pi )^d }
 \delta ( \mu - \epsilon_{\bd{k}} )
 \end{equation}
is the density of states at the chemical potential.
\end{appendix}

\end{document}